\documentclass[11pt]{article}

\usepackage{times}
\usepackage{makecell}
\usepackage{hyperref}
\usepackage{xcolor}		% make links dark blue
\definecolor{darkblue}{rgb}{0, 0, 0.5}
\hypersetup{colorlinks=true,citecolor=darkblue, linkcolor=darkblue, urlcolor=darkblue}
\usepackage[a4paper, total={6in, 8in}]{geometry}
\usepackage{amsmath}
\usepackage{graphicx}
\usepackage{tikz}
\usepackage{amssymb}
\usetikzlibrary{arrows}
\usepackage{amsthm}
\usepackage{float}
\usepackage{mathrsfs}
  \usepackage{setspace}
\usepackage[final]{pdfpages}
\usepackage[normalem]{ulem}
\usepackage{authblk}

\usepackage[super,round]{natbib}

\makeatletter

      \theoremstyle{plain}

       \newtheorem{lemma}{Lemma}
              \newtheorem{corollary}{Corollary}

       \newtheorem{definition}{Definition}
\newcommand{\expit}{\text{expit}}
\newcommand{\logit}{\text{logit}}
\newcommand\@erelb@r[1]{%
  \mathrel{\tikz[baseline=-.5ex]\draw[#1] (0,0)--(0.3,0);}
}

\newcommand{\erelbar}[1]{\@erelbar#1}
\def\@erelbar#1#2{%
  \ifcase\numexpr#1*4+#2\relax
    \@erelb@r{-}\or     % 00
    \@erelb@r{->}\or    % 01
    \@erelb@r{-|}\or    % 02
    \@erelb@r{->|}\or   % 03
    \@erelb@r{<-}\or    % 10
    \@erelb@r{<->}\or   % 11
    \@erelb@r{<-|}\or   % 12
    \@erelb@r{<->}\or   % 13
    \@erelb@r{|-}\or    % 20
    \@erelb@r{|->}\or   % 21
    \@erelb@r{|-|}\or   % 22
    \@erelb@r{|<->|}\or % 23
    \@erelb@r{|<-}\or   % 30
    \@erelb@r{|<->}\or  % 31
    \@erelb@r{|<-|}\or  % 32
    \@erelb@r{|<->|}    % 33
  \else
    \@wrong
  \fi
}
\makeatother

\newcommand{\gol}{
\overset{\mathcal{L}}{\to}
}

\title{Relative Sparsity for Medical Decision Problems}
\author[1,2]{Samuel J. Weisenthal}
\author[1]{Sally W. Thurston}
\author[1]{Ashkan Ertefaie}
\affil[1]{Department of Biostatistics and Computational Biology, University of Rochester School of Medicine and Dentistry, Rochester, NY}
\affil[2]{Medical Scientist Training Program, University of Rochester School of Medicine and Dentistry, Rochester, NY}
\affil[ ]{\textit {\{Samuel\_Weisenthal,\ 
Sally\_Thurston,\
Ashkan\_Ertefaie\}@URMC.Rochester.edu}}

\begin{document}
\maketitle

%\normalsize
%\tableofcontents

\section*{Abstract}
Existing statistical methods can estimate a policy, or a mapping from covariates to decisions, which can then instruct decision makers (e.g., whether to administer hypotension treatment based on covariates blood pressure and heart rate).
There is great interest in using such data-driven policies in healthcare.
 However, it is often important to explain to the healthcare provider, and to the patient, how a new policy differs from the current standard of care.   
This end is facilitated if one can pinpoint the aspects of the policy (i.e., the parameters  for blood pressure and heart rate) that change when moving from the standard of care to the new, suggested policy. 
To this end, we adapt ideas from Trust Region Policy Optimization (TRPO). In our work, however, unlike in TRPO, the  difference between the suggested policy and standard of care is required to be sparse, aiding with interpretability.  
This yields  ``relative sparsity," where, as a function of a tuning parameter, $\lambda$, we can approximately control the number of parameters in our suggested policy that differ from their counterparts in the standard of care (e.g., heart rate only).  We propose a criterion for selecting $\lambda$, perform  simulations, and illustrate our method with a real, observational  healthcare dataset, deriving a policy that is easy to explain in the context of the current standard of care. Our work promotes the adoption of data-driven decision aids,  which have great potential to improve health outcomes.

\newcommand{\realabs}{For example, in the setting of hypotension, one might be interested in deciding, based on a patient's blood pressure and heart rate, whether to administer a medication that increases blood pressure, such as a vasopressor.}

%\keywords{{Lasso, Trust Region Policy Optimization, Individualized Medicine, Causal Inference, Reinforcement Learning}}

% \jnlcitation{\cname{%
% \author{},
% \author{},
% \author{},
% \author{}, 
% \author{}} (\cyear{}),
% \ctitle{}, \cjournal{}, \cvol{}.}

\maketitle

% \footnotetext{\textbf{Abbreviations:} ANA, anti-nuclear antibodies; APC, antigen-presenting cells; IRF, interferon regulatory factor}

\section{Introduction}
\label{sec:intro}
Although risk models for mortality and morbidity are commonly used in the healthcare setting, decision models, which can provide guidance with respect to which  treatment to choose,  are not. Much in the way that risk models can help providers and patients  determine prognosis, decision models can help providers and patients make better, or even optimal, treatment decisions.  There has therefore been great interest in the statistics and machine learning communities on developing methodology to estimate decision models from data. \citep{Chakraborty2013,murphy2003optimal,futoma2020popcorn}  For example, authors have developed methods with applications to the management of hypotension, \citep{futoma2020popcorn,gottesman2020interpretable} sepsis, \citep{raghu2017continuous} and diabetes. \citep{ertefaie2018constructing,luckett2019estimating}  However,  there remains a major barrier to widespread adoption of decision models: it is sometimes difficult to justify to decision-makers why they should replace the established standard of care, or their ``behavioral policy,'' with a new, suggested policy.  In this work, we provide methodology for deriving policies that are easy to explain in the context of the standard of care, and whose adoption is therefore easy to justify.

To develop policies that are easy to explain in the context of the standard of care, we build on the policy search framework, \citep{ng2000algorithms,sutton2018reinforcement} in which one defines a reward function and attempts to find a policy that will maximize it directly.  In order to perform policy search, one uses importance sampling \citep{precup2000eligibility,thomas2015safe} or policy gradients ,\citep{sutton2018reinforcement} which are equivalent to certain techniques in inverse probability weighting. \citep{horvitz1952generalization,robins1994estimation,Chakraborty2013} Policy search differs from Q-learning and model based reinforcement learning, where the value function and transition probabilities are modeled, respectively. \citep{watkins1992q, schulte2014q, ertefaie2021robust}  In policy search, one specifies a model only for the policy, and the expected reward is optimized directly as a function of the policy parameters.  This is convenient when one wishes to place a constraint on the policy itself, as we will  in the current work.  
%Often in policy search, it is important to include some constraint, because the variance of policy search estimators can be high \citep{thomas2015safe,Chakraborty2013}, but we do so for a different reason: relative interpretability.

Our work is related to Trust Region Policy Optimization (TRPO), \citep{schulman2015trust} where one maximizes some reward,  but also requires that the suggested policy be in some sense close to the data-generating behavioral policy. In the current literature, TRPO is primarily applied to non-medical problems, such as updating a robot's  behavior without taking too large of a step. In the robotics setting, since one can simply change the code by which a machine behaves, there is no need to justify a change in policies.     Hence, in  TRPO, 
%the constraint is not on the difference between the parameters of the suggested policy and behavioral policies (which, we believe, is useful for justifying the suggested policy), but rather on the difference between the outputs (actions) generated by the two policies. As a consequence, in Trust Region Policy Optimization}, 
the difference between the suggested and behavioral policies is a \textit{black box}; there is no requirement that the policies be parameterized, and if they are, there is no guarantee that the difference between the parameters of the two policies be interpretable.
%This is appropriate in the robotics setting, where one can simply change the code by which a machine behaves. 
 In healthcare, however, it is important to explain to patients and providers why they should shift from their current policy, which is the standard of care, to a new policy. To this end, our main methodological development is a  \textit{relative sparsity} penalty on the parameters of the suggested  and behavioral policies. We aim to provide a suggested policy such that there is a sparse, and therefore an \textit{interpretable} difference between the parameters of the suggested and behavioral policies, facilitating the explanation and justification of the suggested policy. 
 
  Interpretability is a widely discussed topic in statistical modeling in healthcare. \citep{lipton2018mythos,rudin2019stop}
  Our focus here is  not on the standard notion of the interpretability of a model,  but instead on the interpretability of the difference between two models. Since  sparsity is thought to improve interpretability by reducing the cognitive load on the end-user, \citep{miller2019explanation,du2019techniques,yao2022policy} we build on the sparsity-inducing Lasso penalty \citep{tibshirani2011regression} and its extensions to sparse reinforcement learning. \citep{yang2019regularized, song2015penalized, kolter2009regularization, qin2014sparse, liu2012regularized, hao2020sparse, liu2019regularization, hoffman2011regularized}    In contrast to these studies, where the Lasso constraint region is centered at zero, our Lasso constraint region is {\it centered} at the parameters of the behavioral policy, giving us a type of relative Lasso, which leads to sparse differences between the parameters of the suggested and behavioral policies. {   One can view this recentered Lasso as a nonstandard case of  a fusion penalty, where the behavioral policy parameter is constrained to equal its maximum likelihood estimator (for a description of fusion penalties, see e.g. \cite{price2014fusion}).  In other words, unlike in the standard formulation of fusion penalties,\cite{tibshirani2005sparsity,ding2022cooperative} we do not jointly shrink the suggested policy parameter toward the behavioral policy parameter and the behavioral policy parameter toward the suggested policy parameter (this would involve a joint optimization over both the behavioral and suggested policy parameters). Instead, as in a recentered Lasso, we only shrink the suggested policy parameter toward the behavioral policy parameter. We do so with a two-stage approach: first we estimate the behavioral policy, and then we estimate the suggested policy, making use of the estimate of the behavioral policy. This two-stage approach is key to the decision making application, because the estimate of the behavioral policy cannot be biased in the direction of another parameter, which would occur in a typical fusion.} %For example, note that fused lasso\cite{tibshirani2005sparsity} encourages sparsity  in the coefficient differences. {However, here, our goal is not to fuse the coefficients of the suggested policy but to pull the coefficients of the suggested policy toward the corresponding coefficients in the behavioral policy,} which is done by recentering the Lasso. }  
Although our approach is different from the {standard} Lasso, our focus on a Lasso-type penalty, which constrains the parameters %instead of the outputs (actions) 
of the two policies, serves to distinguish our work not just from Schulman et al., \citep{schulman2015trust} but also from work on contrastive interpretability in reinforcement learning. \citep{puiutta2020explainable} 
A recent example of contrastive interpretability research is Yao et al., \citep{yao2022policy} which employs a penalty similar to TRPO, \citep{schulman2015trust} and although Yao et al. \citep{yao2022policy} provides a sparse list of the actions at which two policies differ,  the difference between the parameters of the policies in Yao et al. \citep{yao2022policy} is still, as in Schulman et al., \citep{schulman2015trust} a (non-sparse) black box.
%(note  that we were led to Schulman et al. \citep{schulman2015trust} by way of the closely connected paper Futoma et al. \citep{futoma2020popcorn}, which indirectly inspired much of our work and was written by the same group that wrote Yao et al. \citep{yao2022policy}).

 In our work, however, in contrast to both Yao et al. \citep{yao2022policy} and Schulman et al., \citep{schulman2015trust} we generate a sparse difference between the \textit{parameters} of the suggested and behavioral policies, which gives a succinct explanation as to \textit{why}, at the level of the weights placed on different covariates, the two policies might disagree.
% Although the methodology in Yao et al. \citep{yao2022policy} is different from our own, both the method in Yao et al. \citep{yao2022policy} and the method that we propose  employ sparsity to reduce the cognitive load on the end-user, which is believed to be an effective strategy \citep{miller2019explanation,du2019techniques}. 
 %That we share this goal  reinforces its importance, not just  for medicine, which is the motivating setting in the work we describe here, but also for the reinforcement learning  in general.
%We develop our methodology for an adaptive Lasso \citep{zou2006adaptive} formulation. \citep{cox1975note,leamer1974false}.
Ultimately, our proposed  methodology allows for the derivation of a policy that is easy to explain in the context of the standard of care.  In addition to proposing a new objective function, we develop a problem-specific criterion for selecting the regularization parameter, $\lambda$, which  dictates the tradeoff between expected reward, i.e., value, and relative sparsity.  Our work facilitates the justification and adoption of data-driven treatment strategies,  and ultimately enhances our ability to translate decision aids into the clinic, where they might substantially improve health outcomes.

\section{Data and framework}
\label{backgroundmdp}
Consider the single stage decision problem, which is a special case of a Markov Decision Process, \citep{bellman1957markovian} and encompasses many problems in medicine.  Let us have an initial state, $S_0\in \mathbb{R}^K$, which is comprised of $K$ covariates.  The dimension $k$ of the state is denoted $S_{0,k}$, and each dimension corresponds to a different covariate. One covariate  may be, e.g., a patient's blood pressure, and another might be heart rate.  Let us also have a binary action, $A_0\in\{0,1\},$ which may be, e.g., the administration of a medication, such as a vasopressor, which constricts the vasculature, and is sometimes used in the setting of low blood pressure, or hypotension.  Let the final state be $S_1.$   Let us observe $i=1,\dots,n$ independent and identically distributed  trajectories of the form $\left\{S_{i,0},A_{i,0},S_{i,1}\right\},$ where each trajectory corresponds to one patient.   A trajectory is sampled from a true distribution, denoted by $P_0(S_0,A_0,S_1),$
 which can be factored into an initial state distribution, $P_0(S_0)$, a  transition probability, $P_0(S_1|A_0,S_0),$ and an action distribution,  $P_0(A_0|S_0).$   We will denote an arbitrary action distribution by dropping the subscript 0.  In other words, an arbitrary action distribution will be denoted $P(A_0|S_0),$ and we will call this arbitrary action distribution a ``policy" and refer to it from now on as $\pi(A_0|S_0)$ as is convention. \citep{sutton2018reinforcement} 
 
 Define a deterministic reward function, 
 %\begin{equation}
 %\label{eq:R}
$R(S_0,A_0,S_{1}),$
 %\end{equation} 
 which may be, e.g., the patient's final blood pressure, which we may want to maximize if the patient is initially hypotensive.
 In reinforcement learning, dynamic treatment regimes, and control theory, \citep{Chakraborty2013, sutton2018reinforcement, puterman2014markov} we often seek a policy that will give us trajectories with higher reward.   For example, one can alter the policy by which one assigns vasopressors to obtain better final blood pressures. 
 %Better characterizing how the policy should change can help the provider and patient make better decisions.
 We will categorize policies according to the following definition.
 \begin{definition}
 \label{def:determStoch}
 A policy $\pi(A_0=a_0|S_0=s_0)$ is deterministic if $\pi(A_0=a_0|S_0=s_0)=0$ or $1$ for all $a_0,s_0.$  A policy $\pi(A_0=a_0|S_0=s_0)$ is stochastic if $0<\pi(A_0=a_0|S_0=s_0)<1$ for all $a_0,s_0.$
 \end{definition}
 In our work, we will  be targeting a stochastic policy, which turns out to be essential to achieving our goal of deriving a policy that is justifiable with respect to the current standard of care, as discussed in Section \ref{sect:l0l1}.  
 We will further parameterize the policy with  $\beta=(\beta_{1},\dots,\beta_{K})^T$, so that
\begin{align}
\label{eq:thetapolicy}
&\pi_{\beta}(A_0=1|S_0=s_0)=\expit(\beta^T s_0)=
\frac{\exp({\beta^Ts_0})}{1+\exp({\beta^Ts_0})}.
\end{align}
For interpretability, in this study, we will only consider $\beta^Ts_0,$ which is linear in the parameters $\beta$, and we will not consider basis  expansions ---see, e.g., Hastie et al. \citep{hastie2009elements} for more discussion of basis expansions--- of $S_0$ (i.e., $S_0^2, S_0^3,\dots$).  

Since $A_0$ is binary, we have that \[\pi_{\beta}(A_0=a_0|S_0=s_0)= \pi_{\beta}(A_0=1|S_0=s_0)^{a_0}(1-\pi_{\beta}(A_0=1|S_0=s_0))^{1-a_0}.\] 
We will parameterize the behavioral policy with  $b=(b_{1},\dots,b_{K})^T,$ so that 
\begin{align}
\label{eq:bpolicy}
\pi_{b}(A_0=1|S_0=s_0)=\expit(b^T s_0)=
\frac{\exp({b^Ts_0})}{1+\exp({b^Ts_0})}.
\end{align}
Let  the parameter of the true behavioral policy be $b_0$, so that the data is truly sampled from the distribution \[P_0(S_0,A_0,S_1)=P_0(S_0)\pi_{b_0}(A_0|S_0)P(S_1|A_0,S_0).\]
The likelihood of a trajectory under the true initial state distribution, the true transition distribution, and an arbitrary policy, $\pi_{b},$  is $P_{b}(S_{0},A_{0},S_{1})=P_0(S_0)\pi_{b}(A_0|S_0)P_0(S_{1}|A_{0},S_{0}).$
Define also
%\begin{equation}
%\label{eq:l}
$l(b) = \log P_{b}(S_{0},A_{0},S_{1}),$  
%\end{equation}
and define the log-likelihood of the observed data as 
\begin{equation}
\label{like}
    l_n(b)=\sum_{i=1}^n\log P_{b}(S_{i,0},A_{i,0},S_{i,1}),
\end{equation}
where here and elsewhere an ``$n$'' subscript denotes an estimator from data with $n$ observations.
If we consider the maximizer of the log-likelihood,
%\begin{equation}
%\label{eq:bn}
$b_n=\arg\max_{b}l_n(b),$
%\end{equation}
then $b_n$ is an estimator for $b_0,$ and $\pi_{b_n}$ is an estimator for $\pi_{b_0}.$ 
%Note that $\pi^{\mathscr{B}}$ might refer to a completely nonparametric estimator.  A parametric version would be $\pi_b^{\mathscr{B}}.$  Soon, we will specialize to the parametric behavioral policy case, and then we will cease to write $\pi^{\mathscr{B}}_b$ and just write $\pi_b.$

 Much of the existing literature on dynamic treatment regimes and reinforcement learning focuses on finding a policy, $\pi_{\beta},$ that maximizes the expected reward, which  is conventionally called ``value.'' \citep{sutton2018reinforcement}
 For any arbitrary parameter $\beta,$ since $A_0\in\{0,1\}$ and $S_0,S_1\in\mathbb{R}^K,$ define the value, $V_0(\beta),$ under a policy, $\pi_{\beta},$ as the expected reward,
 \begin{equation}
 \label{eq:Etheta}
V_0(\beta)= E_{\beta}R(S_0,A_0,S_1)=\int_{s_0}\sum_{a_0}\int_{s_1}R(s_0,a_0,s_1)p(s_1|a_0,s_0)\pi_{\beta}(a_0|s_0)p(s_0){ds_{1}ds_{0}}.
 \end{equation}
 In our work, it is not our goal to maximize $V_0$ alone, but $V_0$ will be a part of our objective function, so we will now describe the properties of its maximizer.   If we define $\pi_0=\arg\max_{\pi}V_0(\pi)$ to be the nonparametric policy that maximizes $V_0,$ then it can be shown, {as in existing literature,} \citep{puterman2014markov,Chakraborty2013,jones2022valid} that
\begin{equation}
\label{eq:determinism}
\pi_{0}(A_0=1|s_0) = I(E( R|S_0=s_0,A_0=1)-E(R|S_0=s_0,A_0=0)>0),
\end{equation}
where we suppress the arguments of $R$ for compactness, and $I(\cdot)$ is an indicator function. {Although it is a known result, we show that Equation (\ref{eq:determinism}) holds using our notation, and for our problem setting, in Appendix \ref{app:indicator}.}
\newcommand{\indicator}{
Write
\begin{align*}
  V_{0}(\pi)
    &= E_{\pi}\left( R(S_0,A_0,S_{1})\right)  \\
  &= E_{\pi}\left(E( R(S_0,A_0,S_{1})|S_0,A_0)\right)  \\
     %&=\int_{s_0}\sum_{a_0}E\left( R|S_0=s_0,A_0=a_0\right)\pi(A_0=a_0|s_0)p(s_0)d{s_{0}} \\
          &=\int_{s_0}E\left( R|S_0=s_0,A_0=1\right)\pi(A_0=1|s_0)p(s_0)d{s_{0}}\\
          &+\int_{s_0}E\left( R|S_0=s_0,A_0=0\right)\pi(A_0=0|s_0)p(s_0)d{s_{0}} \\
          &=\int_{s_0}E\left( R|S_0=s_0,A_0=1\right)\pi(A_0=1|s_0)p(s_0)d{s_{0}}\\
          &+\int_{s_0}E\left( R|S_0=s_0,A_0=0\right)(1-\pi(A_0=1|s_0))p(s_0)d{s_{0}} \\
         %&=\int_{s_0}E\left( R|S_0=s_0,A_0=1\right)\pi(A_0=1|s_0)p(s_0)d{s_{0}}\\
          %&-\int_{s_0}E\left( R|S_0=s_0,A_0=0\right)\pi(A_0=1|s_0)p(s_0)d{s_{0}}\\
          %&+\int_{s_0}E\left( R|S_0=s_0,A_0=0\right)p(s_0)d{s_{0}} \\
          &=\int_{s_0}\left(E\left( R|S_0=s_0,A_0=1\right)-E\left( R|S_0=s_0,A_0=0\right)\right)\pi(A_0=1|s_0)p(s_0)d{s_{0}}\\
          &+\int_{s_0}E\left( R|S_0=s_0,A_0=0\right)p(s_0)d{s_{0}} \\
          &=I+II,
\end{align*}
where \[I=\int_{s_0}\left(E( R|S_0=s_0,A_0=1)-E(R|S_0=s_0,A_0=0)\right)\pi(A_0=1|s_0)p(s_0)d{s_0}\] {and}  the second term, $(II),$ does not depend on $\pi,$ so we can ignore it when we consider the maximization of ${V_0}.$
Suppose \[E( R|S_0=s_0,A_0=1)-E(R|S_0=s_0,A_0=0)>0.\] If so, since $\pi(A_0=1|s_0)\leq 1,$ then
\begin{align*}
          &\int_{s_0}\left(E( R|S_0=s_0,A_0=1)-E(R|S_0=s_0,A_0=0)\right)\pi(A_0=1|s_0)p(s_0)d{s_0}\\
  &\leq \int_{s_0}\left(E( R|S_0=s_0,A_0=1)-E(R|S_0=s_0,A_0=0)\right)p(s_0)d{s_0}.
\end{align*}
{Hence, \[\pi_{0}(A_0=1|s_0) = I(E( R|S_0=s_0,A_0=1)-E(R|S_0=s_0,A_0=0)>0).\]}
}
%{ 
%Equation (\ref{eq:determinism}) implies that $\pi_{0}(A_0=1|s_0)=1$ or 0, depending on $s_0.$ Hence, by Definition \ref{def:determStoch}, $\pi_0$ is a ``deterministic'' policy. 
We will discuss the implications of Equation (\ref{eq:determinism}) for our proposed joint objective in Sections \ref{sect:finitebetaolambda} and \ref{sect:l0l1}.
%}

 \section{Importance Sampling}
 \label{caus}

 Ideally,  to estimate $V_0(\beta)=E_{\beta}{R},$  we would just prospectively take actions under the suggested policy, $\pi_{\beta},$ and observe the rewards. However, doing so in  a medical setting is often not possible for ethical reasons, and we can therefore only observe data generated under the behavioral policy, $\pi_{b_0}$.   To therefore estimate the  value that we would obtain had we, possibly contrary to fact, taken actions according to some policy $\pi_{\beta},$ we  need to take a counterfactual expectation. This can be done with importance sampling, \citep{precup2000eligibility,sutton2018reinforcement} which is known in the causal inference literature as inverse probability weighting.
 \citep{horvitz1952generalization,robins1994estimation,Chakraborty2013}   Note further that under assumptions such as positivity, consistency, and no unmeasured confounding, one can show that $V_0(\beta)$ is causally identified. \citep{munoz2012population,ertefaie2018constructing,luckett2019estimating}

We will now describe an estimand for the counterfactual value, $V_0(\beta)$.   Suppose
\begin{equation}
\label{eq:positivity}
0<\pi_{b_0}(A_0=a_0|S_0=s_0)<1\text{ for all }s_0,a_0.
\end{equation}
This assumption is often called positivity, and, by Definition \ref{def:determStoch}, it is equivalent to stochasticity of $\pi_{b_0}$.
%[add Radon Nikodym assumption]}
As in prior work, \citep{precup2000eligibility,thomas2015safe,murphy2001marginal} it can be shown that 
 \begin{equation}
 \label{eq:v0}
 V_{0}(\beta) = E_{b_0}\left\{\frac{\pi_{\beta}(A_0|S_0)}{\pi_{b_0}(A_0|S_0)} R(S_0,A_0,S_1)\right\}.     
 \end{equation}
Equation (\ref{eq:v0}) follows from a density transform and the fact that the initial state and transition distributions cancel, leaving us only  a ratio of the policies (for completeness, a  derivation of Equation (\ref{eq:v0}) is provided in Appendix \ref{vnaughtIPW}).
 %
%{ For completeness, we prove this fact here.}
\newcommand{\vnaughtIPW}{
\begin{align*}
V_{0}(\beta)&=E_{\beta}[R(S_0,A_0,S_1)]\\
&=E_{b_0}\left\{\frac{P_{\beta}(S_0,A_0,S_{1})}{P_0(S_0,A_0,S_{1})} R(S_0,A_0,S_1)\right\}\\
&=E_{b_0}\left\{\frac{P_0(S_0)\pi_{\beta}(A_0|S_0)P_0(S_{1}|A_{0},S_{0})}
{P_0(S_0)\pi_{b_0}(A_0|S_0)P_0(S_{1}|A_{0},S_{0})} R(S_0,A_0,S_1)\right\}\\
&=E_{b_0}\left\{\frac{\pi_{\beta}(A_0|S_0)}{\pi_{b_0}(A_0|S_0)} R(S_0,A_0,S_1)\right\}.
\end{align*}
}
We accordingly define an inverse probability weighted estimator for  $V_0(\beta),$  assuming a known behavioral policy parameter, $b,$ as
\begin{align}
\label{eq:vnest}
V_n(\beta,b)= \frac{1}{n}\sum_{i=1}^n \frac{\pi_{\beta}(A_{i,0}=a_{i,0}|S_{i,0}=s_{i,0})}{\pi_{b}(A_{i,0}=a_{i,0}|S_{i,0}=s_{i,0})} R(S_{i,0},A_{i,0},S_{i,1}).
\end{align}
%Note  that if we let $E$ and $E_n$ be the expectation and the empirical expectation operator, respectively, then \[V_n(\beta,b)=E_n v(\beta,b) \text{ and } V_0(\beta)=Ev(\beta,b)\] where \[v=\frac{\kappa_{\beta}}{\kappa_{b}} G.\]
The parameter of the optimal policy (subject to no constraints, which we will soon change) is 
\begin{equation}
\label{eq:beta0}
\beta_0=\arg\max_{\beta}V_0(\beta).
\end{equation}
We can define the corresponding estimator,  ${\beta_n}=\arg\max_{\beta} V_n(\beta,{b_n}),$ by substituting a plug-in estimator, $b_n,$ for $b_0$.  It is often useful to constrain $\beta$ during this optimization in some way, which will be the focus on the next section.

\section{Trust Region Policy Optimization}

 If we combine the value, $V_n$, with a penalty on the Kullback-Leibler ($KL$) divergence between $\pi_{\beta}$ and some known $\pi_b,$ we recover an off-policy version of the penalized Trust Region Policy Optimization (TRPO) estimator, which is given in Section 4 of Schulman et al., \citep{schulman2015trust} 
\begin{equation}
\label{eq:trust.pol.obj.sch}
\beta_n=\arg\max_{\beta} V_n(\beta,b)-\lambda KL(\pi_{\beta},\pi_{b}).
\end{equation}
The objective function in Equation  (\ref{eq:trust.pol.obj.sch}) is similar to those in Futoma et al. \citep{futoma2020popcorn} and Farahmand et al., \citep{farahmand2016value} since minimizing an empirical version of KL divergence is equivalent to maximizing likelihood (for a proof, see the section on M-estimators in van der Vaart \citep{van2000asymptotic}). Maximizing the objective function defined by Equation (\ref{eq:trust.pol.obj.sch}) yields a $\pi_{\beta_n}$ that ``stays close'' to the  behavioral policy, ${\pi_b}$, which stabilizes the optimization by mitigating instability in the ratio that is present in $V_n$ in Equation (\ref{eq:vnest}).  The objective in Equation (\ref{eq:trust.pol.obj.sch}) is applied predominantly in robotics, where justification of a policy change is not as important, and hence the  difference between $\pi_b$ and $\pi_{\beta_n},$ if they are parameterized at all, which is not required, is not guaranteed to be sparse (i.e., the difference between $b$ and  $\beta_n$ is not guaranteed to be sparse). In a healthcare setting, in contrast, one must convince the healthcare provider and  the patient to adopt a new treatment policy.  Hence, in a healthcare setting, we require that the difference between $\pi_b$ and $\pi_{\beta_n}$  be  \textit{sparse} (i.e., that the difference between  $b$ and $\beta_n$  be sparse).  This sparsity provides relative interpretability, and the ability to justify the suggested policy, which is the deliverable of our proposed method, and which we will describe in the next section.

\section{Methodology}
\subsection{Relative Sparsity}
\label{section:relativeSparsity}

\begin{figure}
    \centering
    \includegraphics[width=0.8\textwidth]{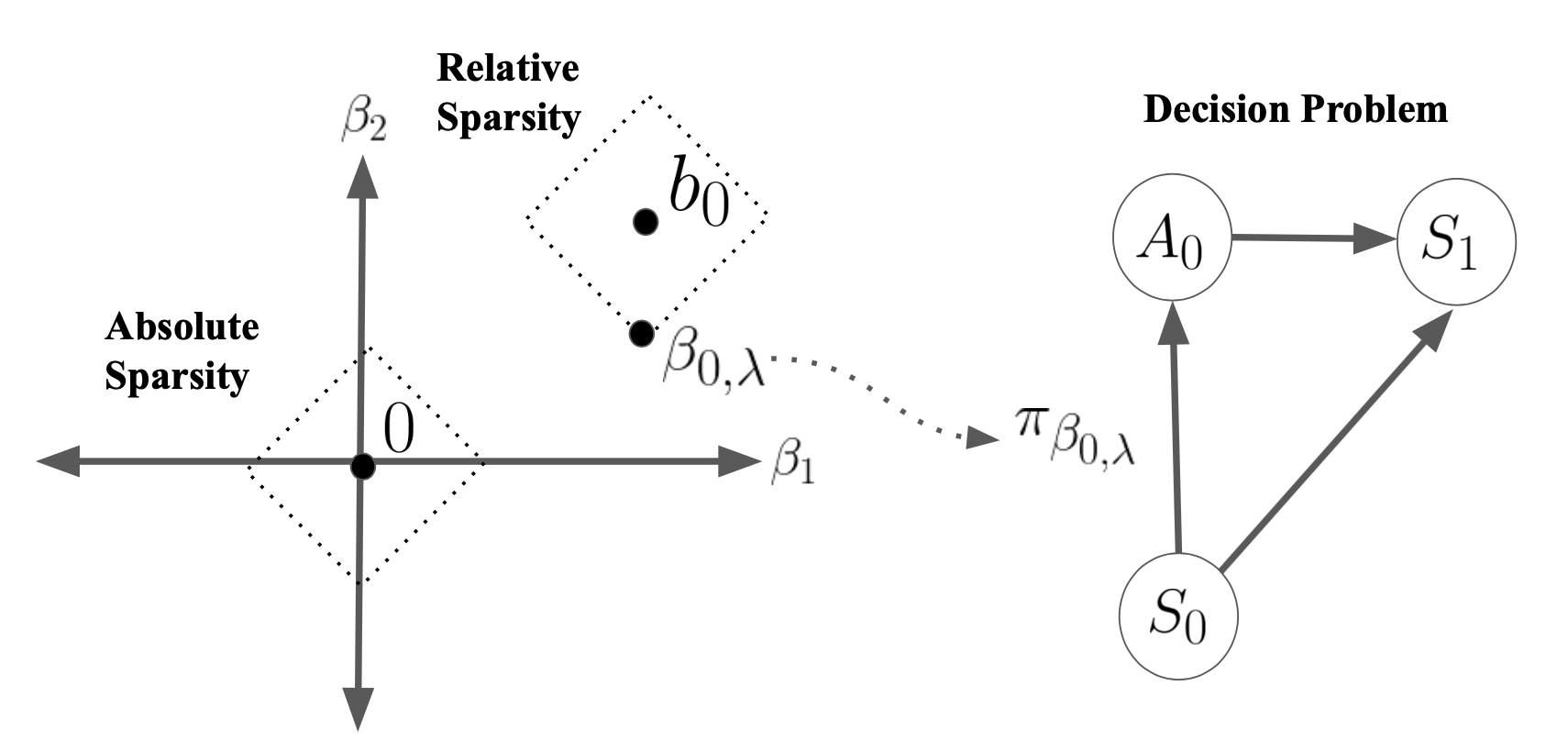}
    \caption{{\bf Constraint region of the relative sparsity penalty in the context of a decision problem.} We show how the relative sparsity penalty recenters the standard diamond shaped constraints of the $L_1$ norm.  If we are optimizing over the parameters  $\beta=(\beta_1,\beta_2)^T$ of a policy  $\pi_{\beta},$ the relative sparsity constraint region is centered at the parameter of the behavioral policy $b_0=(b_{0,1},b_{0,2})^T,$ whereas  the  standard lasso constraint region, which gives absolute sparsity, is centered at the zero vector ${\bf 0}=(0,0)^T$. Our relative sparsity constraint must be considered in the context of a decision problem, shown to the right. The  relative sparsity objective function gives relative interpretability of the policy $\pi_{\beta_{0,\lambda}}$ with respect to the behavioral policy $\pi_{b_0},$ which is not guaranteed under the absolute sparsity constraint. }
    \label{fig:rscon}
\end{figure}

As mentioned, in a healthcare setting, unlike in robotics, we have to convince the provider and patient to adopt a suggested policy. Hence, the difference between $\pi_b$ and $\pi_{\beta_n}$ must be interpretable.  We have already taken one step toward making the difference interpretable by parameterizing our policy.  Our major contribution toward making the difference interpretable, however, is the relative sparsity penalty, which requires that only a few of the coefficients in the suggested policy, $\pi_{\beta_n}(A_0=a_0|S_0=s_0),$ deviate from their respective coefficients in the behavioral policy, $\pi_{b_n}(A_0=a_0|S_0=s_0)$. 
Concretely, our objective is
\begin{align}
\label{eq:j0}
J_0(\beta,b_0,\lambda)=V_0(\beta)-\lambda||\beta-b_{0}||_1.
\end{align}
This penalty yields relative sparsity, which leads to an interpretable difference between the suggested and behavioral policy, and therefore facilitates justification of the suggested policy.  The constraint region implied by this relative Lasso penalty, as it compares to standard Lasso, is shown in Figure \ref{fig:rscon}.

We  define an estimator for $J_0$ to be 
\begin{align}
\label{eq:jn}
J_n(\beta,b_n,\lambda)=V_n(\beta,b_n)-\lambda||\beta-b_{n}||_1.
\end{align}
The true parameters of our relatively sparse policy, $\beta_{0,\lambda}=\arg\max_{\beta} J_0(\beta,b_0,\lambda),$ are estimated using  $\beta_{n,\lambda}=\arg\max_{\beta} J_n(\beta,b_n,\lambda).$ Note that we ensure that each coefficient is penalized equally by scaling the states as described in Appendix \ref{scaling} (assume that any state that we refer to in the simulations, which are in Section \ref{rs:sim}, and the real data analysis, which is in Section \ref{rs:realdata}, is scaled). 

\subsection{A problem-specific $\lambda$ selection criterion}
\label{sec:choosinghyp}

The hyperparameter $\lambda$ controls the tradeoff between relative sparsity and value. In the robotics setting of Schulman et al., \citep{schulman2015trust}  $\lambda$ is often updated iteratively according to Equation (\ref{eq:trust.pol.obj.sch}).  In other words,  a robot perform a task according to policy 1, updates to policy 2 by defining the behavioral policy as policy 1, performs the task according to policy 2, updates to policy 3 by defining the behavioral policy as policy 2, and so on.  In constrast, in healthcare, $\lambda$ must be chosen once, and in a problem-specific manner.

We believe that the appropriate tradeoff between sparsity and value  depends on the decision problem.  For example, consider blood pressure management in the outpatient, primary care setting.  In this setting, we can prescribe a fairly benign blood pressure medication and watch its effect over time. It might be possible to suggest a new policy that changes many of the coefficients with respect to the standard of care, and the providers and patients would still adopt the new policy, because such a policy could  lead to large changes in expected reward, or value, with little risk of doing harm to the patient. %This might allow for a policy that can provide large changes in expected reward, even though it might not be as clear how it changes with respect to the standard of care.  
Now consider the inpatient, intensive care unit setting, where stakes are higher.  Blood pressure control might impact the patient's mortality within minutes. In this  setting, a policy must be clearly justifiable relative to established care practices.  It might be better if very few of the coefficients  diverge from the standard of care, and small increases in expected reward, or value, might be preferable to larger increases that risk causing  harm to the patient.  

\begin{figure}
    \centering
    \includegraphics[width=0.8\textwidth]{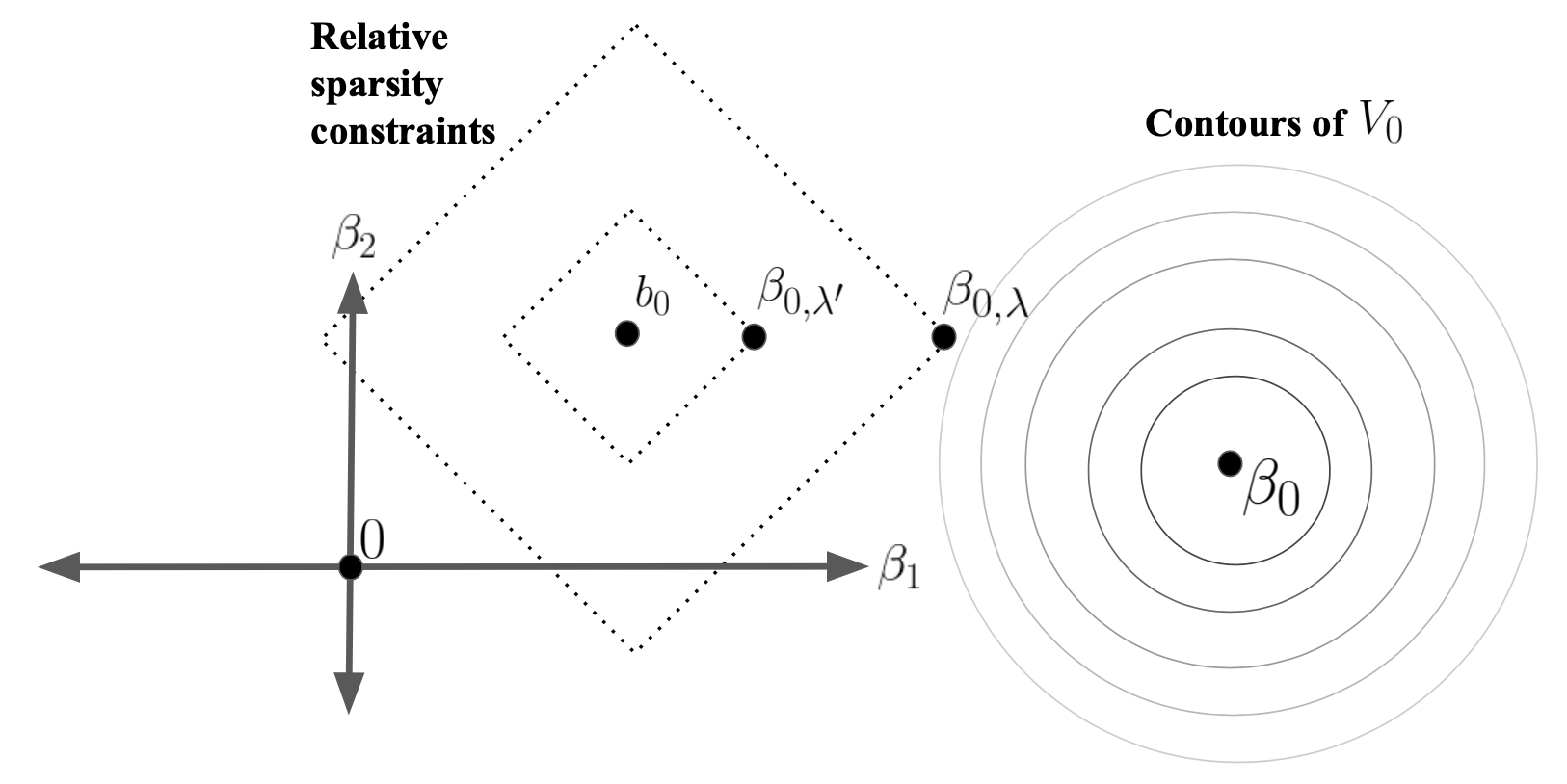}
    \caption{{\bf Relationship between $V_0({\beta_{\lambda}})$ and $V_0({\beta_{\lambda'}})$ when $\lambda'{>}\lambda$.} Consider, as in Figure \ref{fig:rscon}, a bivariate parameter $\beta=(\beta_1,\beta_2)^T,$ which will be chosen to optimize the joint objective, $J_0(\beta,b_0,\lambda),$  in Equation (\ref{eq:j0}). Note that $\beta_0=\arg\max_{\beta}V_0$ (also, $\beta_0=\arg\max_{\beta}J_0(\beta,b_0,0)$) is shown beyond the horizontal axis to indicate that, without loss of generality, $\beta_{0,1} $ happens to be arbitrarily large, and that the contours around $\beta_0$ correspond to the contours of $V_0(\beta)$, where larger diameter of the contours indicates smaller $V_0(\beta)$. Also shown are two possible solutions for $\beta,$ $\beta_{0,\lambda'}$ and $\beta_{0,\lambda},$ under relative penalties $\lambda'$ and $\lambda$, respectively. Since $\lambda' {>} \lambda$, we see that the diamond constraint of the $L_1$ norm for the relative sparsity condition is larger for $\lambda$ than for $\lambda'$.  Since, for a given $\lambda,$ the maximizer of the joint objective, $J_0,$ must be as close as possible to $\beta_0$ but  also must fall within the $\lambda$-specific diamond level set, we have that $\beta_{0,\lambda}$ will be closer than $\beta_{0,\lambda'}$ to $\beta_0$. Since  $\pi_{\beta_0}$  maximizes $V_0$, we have therefore that $V_0({\beta_{0,\lambda}})\geq V_0({\beta_{0,\lambda'}}).$}
    \label{fig:lambdalambdaprime}
\end{figure}

We can obtain  policies for these two different scenarios with different settings of $\lambda$, since there will be a tradeoff with respect to expected reward, or value, and relative sparsity.  Ideally, one would choose a policy with a difference from behavior that is just sparse enough, but not more sparse, since relative sparsity decreases  value.  
Formally, {let us choose a policy with value that is greater than $V^{min}$, which we might define to be the minimal clinically acceptable value. To set $V^{min},$ one might consult guidelines,\cite{demers2021physiology} or, in a more data-driven fashion, one might set $V^{min}$ to be a value that is some number of standard errors above the value of the behavioral policy.}  If we determine that, in addition, based on the nature of the decision problem, we would like only approximately $C$ coefficients to differ from behavior (this is perhaps a crude way to measure the ``stakes,'' but it is at least quantitative), then, if $I(\cdot)$ is an indicator function, we 
target
\begin{align}
\label{eq:lambda.crit.diff}
\lambda_0
= {\max}
\left\{
\arg\min_{\lambda: {V_{0,\lambda}\geq V^{min}}} |D_{0,\lambda}-C|
\right\},
\end{align}
where \[D_{0,\lambda}=\sum_{k=1}^K I\left(
|\beta_{0,\lambda,k}-b_{0,k}|>\Delta
\right)\]
is the number of coefficients that  diverge beyond some tolerance $\Delta\in \mathbb{R}$ from their behavioral counterparts. Hence, we are targeting a sparse policy within the set of policies that have value greater than $V^{min}$.
%Although $V^{min}$ can be set based on clinical background, we also propose some objective choices for $V^{min}$, such as  setting $V^{min}$ to a value that achieves at least one standard error above the behavioral policy value {\color{green}, since adding a standard error gives us an upper quantile of the distribution of $V_n$}) or referring to a clinical target (e.g., based on the Surviving Sepsis Guidelines \cite{dellinger2004surviving}).
We further take a {maximum} in Equation (\ref{eq:lambda.crit.diff}) to ensure that, of all the policies with $D_{0,\lambda}=C$ { and value greater than $V^{min},$} we take the one { that has coefficients that are as close as possible to the coefficients of the behavioral policy. Note that one could replace the $\max$ in Equation (\ref{eq:lambda.crit.diff}) with a $\min$, which might yield a policy with value that overshoots $V^{min}.$ However, in doing so, we might
lose closeness to the behavioral policy. Since we are guaranteed value above $V^{min}$, we recommend instead using a $\min$ in (\ref{eq:lambda.crit.diff}) and increasing $V^{min}$ if a policy with higher value is desired.} { Closeness to behavior, which translates to closeness of the suggested policy coefficients to the behavioral policy coefficients, leads to a more clinically palpable policy, which promotes adoption.}
%, even if we sacrifice some value (we could have instead taken a $\max$ in (\ref{eq:lambda.crit.diff}), which would have given us the policy with the highest value in the set of policies with the desired level of relative sparsity, but, because we can set $V^{min}$ to be larger if we would like more value, we instead choose to target closeness to behavior in this part of the equation, which turned out to be important in the real data analysis).} 
If we look at an example of the level sets around $\beta_0$ and $b_0$, which we show in Figure \ref{fig:lambdalambdaprime}, we {can see that, generally,}
\begin{equation}
\label{eq:lamsmallHighVal}
\lambda\leq \lambda'\implies V_0({\beta_{0,\lambda}})\geq V_0({\beta_{0,\lambda'}}),
\end{equation}
{and, hence, we lose some value by taking the maximum in (\ref{eq:lambda.crit.diff}), but we gain the maximum closeness to behavior within the set of policies that have value of at least $V^{min}$.}

We acknowledge that the user-defined number of diverging covariates, $C$, might be challenging to set, but we still give practitioners the option; if there is no clear best choice of $C,$ it can be simply set to 1 to target maximum sparsity within the class of policies that achieve value greater than or equal to $V^{min}$. In other words, we are choosing a $\lambda_0$ that gives a policy {with value above some threshold, $V^{min}$, but that has} $D_{0,\lambda_0}$  parameters that diverge from behavior, and we want $D_{0,\lambda_0}$ to be as close as possible to  $C$.  Note that $D_{0,\lambda}$ depends on $\lambda$ through $\beta_{0,\lambda}$, which maximizes $J_0$ in Equation (\ref{eq:j0}). Note also that $\Delta $  can be set according to the magnitudes of the coefficients; e.g., if the coefficients in the training data are large, we can set $\Delta\approx 1$, whereas if the coefficients are small, we can set $\Delta\approx 0.01$. 

We estimate $\lambda_0$ with
\begin{align}
%\label{eq:lambda.crit.diff}
\lambda_n
= {\max}
\left\{
\arg\min_{\lambda: { V_{n,\lambda}\geq V^{min}}} |D_{n,\lambda}-C|
\right\},
\end{align}
where \[D_{n,\lambda}=\sum_{k=1}^K I\left(
|\beta_{n,\lambda,k}-b_{n,k}|>\Delta
\right)\]
is the number of coefficients that diverge  empirically from their behavioral counterparts.
 Since $D_{n,\lambda_n}$ is random, taking values $d_{n,\lambda_n},$  we will show tables of its empirical distribution when we perform simulations in Section \ref{rs:sim}.

To aid in choosing $C$ and $\Delta$, we can perform a visualization of the coefficients in the training set as we vary $\lambda$. 
%We then choose $\lambda$ by { minimizing Equation (\ref{eq:lambda.crit.diff})}.  
As long as only the training data is used, this procedure can be interactive; it may be the case that for some problems we cannot obtain a policy with approximately $C$ coefficients that diverge {that also has value above $V^{min}$}, in which case we can assess the results in the training data and, in collaboration with the decision makers, perhaps choose a larger target $C$. {When $V^{min}$ is large, this may require decision makers to reduce their expectations for sparsity, and vice versa.} Having set {$V^{min}$ }, $C,$ and $\Delta$, we then choose $\lambda$ by  minimizing Equation (\ref{eq:lambda.crit.diff}).  One can then use the test data to estimate the value of the final policy, as we show in Sections \ref{rs:sim} and \ref{rs:realdata}. {We provide pseudocode in Appendix \ref{pseudocode}.}

%\section{{ The behavior of the penalized estimator $\beta_{n,\lambda}$}}
%\label{sect:infinitebetaolambda}
\begin{corollary}
\label{coro:deterImpInf}
Recall Definition \ref{def:determStoch}, which states that a policy $\pi$ is deterministic if and only if $\pi(a_0|s_0)=1$ or $0$ for all $a_0, s_0.$  Assume the  form of the optimal policy follows Equation (\ref{eq:thetapolicy}), i.e., $\pi_{\beta_0}=\expit(\beta_0^Ts_0),$ and that $S_{0,k}<\infty$ for all $k=1,\dots,K$.   A policy $\pi_{\beta_0}$ is deterministic if and only if one of the entries in its parameter vector $\beta_0$ is infinite (or becomes arbitrarily large in magnitude). 
\end{corollary}
Corollary \ref{coro:deterImpInf}, for which we provide a formal proof in Appendix \ref{app:deterImpInf}, follows from the fact that the $\expit$ function equals one or zero only when its input is positive or negative infinity (or arbitrarily large).
%\begin{proof}
\newcommand{\determImpliesInf}{
Recall, by Definition \ref{def:determStoch}, that if a policy $\pi$ is deterministic, then $\pi(A_0=1|S_0=s_0)=1$ or $0.$ Recall that by Equation (\ref{eq:thetapolicy}), \[\pi_{\beta_0}(A_0=1|S_0=s_0) = \expit(\beta_0^Ts_0).\]
Now, if $\logit(p)=1/(1-p),$ then 
\begin{align*} 
&\expit(\beta_0^Ts_0)=1 \iff \logit(\expit(\beta_0^Ts_0))=\logit(1)
\iff\beta_0^Ts_0=\infty \iff \exists\ k\ \text{ s.t.}\ |\beta_{0,k}| =\infty,
\end{align*}
since $s_{0,k}<\infty,\ \forall k$ and $\logit(p)\rightarrow\infty$ as $p \rightarrow 1$. 

We say however that the entries of $\beta_0$ tend in magnitude toward infinity rather than ``equal'' infinity in magnitude because, for some rewards, setting entries of $\beta_0$ equal in magnitude to infinity can lead to an undefined linear predictor, $\beta_0^Ts_0$. For example, let $S_0\in \mathbb{R}^2$ and consider the maximization of $V_0=E_{\beta}R$ with $R(S_0,A_0,S_1)=(S_{0,1}-2S_{0,2})A_0$. 
Note that
\begin{align*}
E_{\beta}R(S_0,A_0,S_1)&=E_{\beta}[(S_{0,1}-2S_{0,2})A_0]\\
&=E_{\beta}[E[(S_{0,1}-2S_{0,2})A_0|S_0]]\\
&=E_{\beta}[(S_{0,1}-2S_{0,2})E[A_0|S_0]]\\
&=E_{\beta}[(S_{0,1}-2S_{0,2})\expit(\beta^T S_0)]\\
&=E_{\beta}[(S_{0,1}-2S_{0,2})\pi_{\beta}(A_0=1|S_0)].
\end{align*}
The last line is maximized by  
\begin{equation}
\label{eq:notInf}
\pi_{\beta_0}=I(S_{0,1}-2S_{0,2}>0).
\end{equation}  
Since we have a policy $\pi_{\beta_0}=\expit(\beta_{0,1}S_{0,1}+\beta_{0,2}S_{0,2}),$
we can approximate this indicator by setting $\beta_{0,1}$ arbitrarily large but finite, and $\beta_{0,2}=-2\beta_{0,1},$ so that 
\[\pi_{\beta_0}=\expit(\beta_{0,1}S_{0,1}-2\beta_{0,1}S_{0,2})=\expit(\beta_{0,1}(S_{0,1}-2S_{0,2})),\] which approximates the indicator in Equation (\ref{eq:notInf}) as $\beta_{0,1}$ becomes large in magnitude.
However, if we set $\beta_{0,1}=\infty,$ we get $\beta_{0,2}=-2\infty=\infty,$ and 
\[\pi_{\beta_0}(A_0=1|S_0=s_0)=\expit(\infty S_{0,1}-\infty S_{0,2}),\]
which is undefined.

More generally, the standard rules of arithmetic must hold for the entries of $\beta_0$, which is not the case when one or more entries of  $\beta_0$ is infinite in magnitude.  Hence, it appears that for some rewards, a linear model might simply not be expressive enough; we can never have $\pi_0$ be equaled by $\pi_{\beta_0}$, since $\pi_0$ is a deterministic indicator function and $\pi_{\beta_0}$ that maximizes the expected reward cannot have coefficients that are infinite in magnitude, and therefore cannot be deterministic. In practice, an expit with large parameters can approximate an indicator function well; i.e., the value $V_0(\beta_0)$ when $\beta_{0,k}$ is arbitrarily large but finite in magnitude for some particular $k$ is essentially the same as if $|\beta_{0,k}|=\infty$.
%\end{proof}
}
We use the terminology ``arbitrarily large in magnitude'' because sometimes parameters do not become infinite in magnitude, but approach infinity in magnitude, such that the policy they parameterize is essentially deterministic. 
Corollary \ref{coro:deterImpInf} will help us interpret our simulation results and give  insight into the behavior of the maximizer of the penalized relative sparsity objective in Equation (\ref{eq:j0}).

\subsection{Stochasticity of $\pi_{\beta_{0,\lambda}}$}
\label{sect:finitebetaolambda}
By Equation (\ref{eq:determinism}), which gives an expression for $\pi_0=\arg\max_{\pi}V_0$ as an indicator function, if $\pi_{\beta_0}$ is assumed to be close to $\pi_0,$ then $\pi_{\beta_0}$  is deterministic. Hence, by Corollary \ref{coro:deterImpInf}, $\pi_{\beta_0}$ will have some  parameters that equal or approach infinity in magnitude.  By the following Lemma, however, $\beta_{0,\lambda}=\arg\max_{\beta}J_0,$ will be finite. 

\begin{lemma}
\label{lemma:finiteBeta0lambda}
For $\lambda>0,$  if we assume Equation (\ref{eq:positivity}), Equation (\ref{eq:thetapolicy}), and that \[E(R(S_0,A_0,S_1)|S_0=s_0,A_0=a_0)<\infty\] for all $s_0,a_0,$ then the policy $\pi_{\beta_{0,\lambda}}$ is stochastic. In other words,  $|\beta_{0,\lambda,k}|<\infty$ for all $k$ (and $\beta_{0,\lambda,k}$ does not become arbitrarily large in magnitude).
\end{lemma}
We provide a proof in Appendix \ref{finiteBeta0lambda}.
%but the general idea is to proceed by contradiction. Suppose that the optimal policy is deterministic, which,  by Corollary \ref{app:deterImpInf}, is the case only if an entry of $\beta_{0,\lambda}$ tends toward infinity.  If $\beta_{0,\lambda}$  tends toward infinity, then the penalty term, $-\lambda||\beta_{0,\lambda}-b_0||,$ tends toward negative infinity, drawing the joint objective in the same direction.  
The finiteness of $\beta_{0,\lambda}$ turns out to play an essential role in the justifiability of $\pi_{\beta_{0,\lambda}}$ with respect to the standard of care, a fact on which we will elaborate  in the next section.

\newcommand{\proofFinite}{
%We prove Lemma \ref{lemma:finiteBeta0lambda}.
%\begin{proof}
{T}o maximize $V_0$ requires, {by  Equation (\ref{eq:determinism})}, that \[\pi_{\beta}(A_0=1|s_0)=I\left(E( R|S_0=s_0,A_0=1)-E(R|S_0=s_0,A_0=0)\right),\]  which only occurs,
by Corollary \ref{coro:deterImpInf}, if $|\beta_{k}|$ tends arbitrarily closely to $\infty$ for some $k.$ However, suppose, by contradiction, that $|\beta_{k}|=\infty$ for some $k$.  In this case, the term $-||\beta-b_0||_1=-\infty,$ since $|b_0|<\infty$ by positivity (Equation (\ref{eq:positivity})) and Corollary \ref{coro:deterImpInf}. In this case, $J_0(\beta)$ cannot be at its maximum. Hence, under the stated assumptions, by Corollary \ref{coro:deterImpInf}, $\pi_{\beta_{0,\lambda}}$ is stochastic.  In general, although $V_0$ draws some entry of $\beta$ in magnitude toward $\infty,$ the penalty draws the same entry in the opposite direction.  
%\end{proof}
}

\subsection{On $L_1$, $L_0$, and absolute sparsity}
\label{sect:l0l1}
%The  parameter $\lambda$ deteremines the degree of stochasticity and the degree of relative sparsity of the policy $\pi_{\beta_{0,\lambda}}$. This stochasticity is a consequence of the $L_1$ penalty. 

{
 {In summary, the key objective of our proposed approach is to facilitate the justification and adoption of data-driven treatment strategies,
and ultimately enhance our ability to translate decision aids into the clinic. This has motivated us to develop a method that simultaneously controls the
(1) closeness, defined as the divergence between the suggested policy probability of treatment, $\pi_{\beta_{n,\lambda_n}}(A_0=1|S_0=s_0)$, and the behavioral policy probability of treatment, $\pi_{b_n}(A_0=1|S_0=s_0)$,  
{and (2) sparsity} between the coefficients of the behavioral policy and the suggested
policy,
 which, we hypothesize,} act in concert to promote the adoption of a new policy in the clinic. {This joint closeness and sparsity can  be achieved by using an $L_1$ norm in the penalty.}
 
Closeness to the standard of care is  important, because it is challenging to specify a reward perfectly, and sometimes therefore it is safer to stay close to the standard of care, which is a vetted guideline in many cases. In our method, the $L_1$ penalty is important because the $L_1$ penalty shrinks, which achieves this closeness to behavior, in addition to selecting, which we need for relative sparsity.  That closeness to behavior is important for safety has been explored extensively in other work (see Achiam's survey \cite{achiam2017constrained} on safe reinforcement learning).  {Note that, in general, closeness to established guidelines is important in a healthcare setting, because dramatic changes are often not easily accepted or may take a very long time to be adopted.\citep{lipton2018mythos,rudin2019stop}}}  

Hence, we use an $L_1$ penalty not for computational reasons, as is commonly the case when an $L_1$ penalty is used to approximate an $L_0$ (best subsets) penalty, but instead because the $L_1$ penalty,  while selecting, which reduces cognitive burden for the end-user, \citep{miller2019explanation,du2019techniques,yao2022policy} also shrinks the coefficients to behavior. 
Hence, the $L_1$ penalty is, for our motivation, as useful in the low-dimensional case as in the high dimensional case. 
%In other words, ({we do not use $L_1$ as a computationally feasible approximation to $L_0,$ as is commonly done, but instead because $L_1$ shrinks when $L_0$ does not}). 
This {desired} shrinkage does not occur with an $L_0$ penalty.  
{Concretely,} define $\beta_{0}^{L_0}$ to be a parameter vector in which the reward-relevant parameters are set to their corresponding entries in $\beta_0=\arg\max_{\beta}V_0$ (also, note that $\beta_0=\arg\max_{\beta}J_0(\beta,b_0,0)$), and the reward-irrelevant parameters are set to their corresponding behavioral counterparts in $b_0$.  This would be the resulting policy if we could  use an $L_0$ rather than $L_1$ penalty, where the $L_0$ policy counts the number of non-behavioral coefficients. The  $L_1$ solution is known to approximate the $L_0$ solution, which is useful because the $L_0$ problem is NP-Hard. \citep{natarajan1995sparse}  The $L_1$ solution does not equal the $L_0$ solution, because the $L_1$ norm shrinks instead of purely selecting coefficients, which is considered a drawback.\citep{bertsimas2016best}  Often, in the literature (e.g., in Tibshirani et al., \citep{tibshirani2011regression}) it is the likelihood that is being penalized, and the selected maximum likelihood estimators are of interest, so  it is  not desirable to shrink (bias) their estimates.

In our case, however, {the fact that the $L_1$ penalty shrinks is a benefit.} If we could exactly obtain the $L_0$ solution, $\beta_0^{L_0}$, at least one of the selected coefficients would tend toward infinity by Corollary \ref{coro:deterImpInf}.  Hence, the  coefficients  that would be selected in the suggested policy would be  arbitrarily larger in magnitude than the unselected coefficients. 
%In our case, however, the unselected coefficients are of interest. The unselected coefficients are the contribution from the behavioral policy.  
Hence,  the suggested policy would  ignore the behavioral covariates, which is not desirable.  The property that we call ``justifiability'' is, in fact, a combination of relative sparsity and also of shrinkage, where the shrinkage must be pronounced enough to allow the behavioral policy coefficients to impact the linear predictor $\beta_{0,\lambda}^Ts_0$ and hence to impact the resulting policy, $\pi_{\beta_{0,\lambda}}$.
Hence, the fact that the $L_1$ norm shrinks is key to providing a suggested policy that is not dominated by the reward-relevant covariates, as would occur with the $L_0$ solution.  Thus, the shrinkage that is commonly viewed as a drawback to the $L_1$ norm is, in our case, crucial to obtaining a policy with the desired properties.

Define also $J_0^{Abs}(\beta,\lambda)$, in contrast to our Equation (\ref{eq:j0}) objective, which is $J_0(\beta,b_0,\lambda)=V_0-\lambda ||\beta-b_0||_1,$ \[J^{Abs}_0(\beta,\lambda)=V_0(\beta)-\lambda||\beta||_1.\] 
Whereas the parameter $\beta_0^{L_0}$  sets the unselected coefficients to their behavioral counterparts, here the parameter ${\beta^{Abs}_0}=\arg\max_{\beta}J_0^{Abs}$ sets these same coefficients  to zero. However, 
because the selected coefficients tend in magnitude toward infinity, $(\beta_0^{Abs})^Ts_0$ effectively equals $(\beta_0^{L_0})^Ts_0,$ and hence $\pi_{\beta_0^{Abs}}$ is indistinguishable from $\pi_{\beta_{0}^{L_0}}$.  Hence, the shrinkage of the $L_1$ penalty is also essential for differentiating the relative sparsity penalty from an absolute sparsity penalty.
{
We show the properties of the relative sparsity objective as it relates to  other reinforcement learning methods in Table \ref{qual}.}
\begin{table}
\tiny
\centering
\begin{tabular}{c|ccccccc}
\hline
             & \thead{Parametric\\ $\pi_{\beta}$} & \thead{Max. \\ value} & \thead{Relative \\sparsity} & \thead{Absolute\\ sparsity} & \thead{Close to\\ behavior} & \thead{Specify\\ behavior} & \thead{Specify value \\or transitions} \\
\hline
Proposed (relative sparsity):\\ $V_n(\beta,b)-\lambda L_1(\beta,b)$ & yes & no & yes & no &  yes & yes &  no \\
\hline
TRPO\cite{schulman2015trust}: \\$V_n(\pi_{\beta},\pi_b)-\lambda KL(\pi_{\beta},\pi_b)$ & no & no & no & no  & yes & yes & no \\
\hline
Absolute relative sparsity: \\$V_n(\beta,b)-\lambda L_0(\beta,b)$ & yes & no & yes & no & no & yes & no \\
%\hline
%Absolute relative sparsity: $V_n(\beta,b) - \lambda L_0(\beta,b)$ & yes & no & no & yes & no & yes & no \\
\hline
Absolute sparsity:\\ $V_n(\beta,b) - \lambda L_0(\beta)$ & yes & no & no & yes & no & yes & no \\
\hline
Absolute sparsity \cite{yang2019regularized}: \\$V_n(\beta,b) - \lambda L_1(\beta)$ & yes & no & no & yes & no & yes & no \\
\hline
Unconstrained\cite{ng2013pegasus}: \\$V_n$ & no & yes & no & no & no  & yes & no \\
\hline
Q-learning\cite{watkins1992q}\\ & no & yes & no & no & no & no  & yes \\
\hline
Sparse Q-learning\cite{song2015penalized}\\ & no & no & no & no & no & no  & yes \\
\hline
Model-based: \\(standard)\cite{sutton2018reinforcement} & no & yes & no & no & no & no & yes \\
\hline
Model-based: \\($V_n-\lambda Likelihood$)\cite{futoma2020popcorn,farahmand2016value} 
& no & yes & no & no & yes & no & yes \\
\hline
\end{tabular}
\caption{ {{\bf The properties of relative sparsity as it relates to other reinforcement learning approaches.} ``Parametric $\pi_{\beta}$'' is whether $\beta$ is parametric. ``Max. value'' indicates whether the policy is optimal with respect to $V_0,$ which will never be the case when there is a penalty.  ``Relative sparsity'' indicates whether there will be only a sparse set of coefficients that differ between the behavioral and suggested policy. ``Absolute sparsity'' indicates whether the policy will have a sparse set of nonzero coefficients, which is equivalent to a relative sparsity to a randomized behavioral policy, which has $b_0=(0,\dots,0)$. ``Close to behavior'' indicates whether the suggested policy probabilities of treatment will be similar to the behavioral policy probabilities of treatment. ``Specify behavior'' indicates whether we must have a model for the behavioral policy; we consider only the policy search ($\arg\max V_n$) objectives for all but the Q-learning and model-based approaches, and hence we consider specification of behavior necessary, because it is necessary for policy search. ``Specify value or transitions'' indicates whether we must propose models for these aspects of the environment and reward structure, which can be non-trivial, and is required in Q-learning and model-based approaches. For the unconstrained approach, we consider the nonparametric optimizer that achieves maximal value. Note that TRPO can be formulated with a parametric policy, $\pi_{\beta}$, but this is not required, and it is often not the case given that the application area is robotics (the policy might be parametric, but it would be over-parameterized as in Deep Learning).}}
\label{qual}
\end{table}

\subsection{{Estimation of the behavioral policy}}
\label{behestim}
 {We now describe estimation of the standard of care (behavioral) policy parameters in more detail. The standard of care policy parameters are estimated as $b_n$ that maximizes the likelihood, $l_n$ (Equation (\ref{like})), of the observed data (this can be estimated by using the generalized linear model (GLM) R package).   The estimate $b_n$  is derived after covariate scaling, so that the covariates for the behavioral policy and for the suggested policy, which are penalized toward one another, pertain to the same data. 

   Note that $b_n$ will be  an estimate of $b_0$, and the
          noise in $b_n$ will impact the estimation of $\beta_{n,\lambda}$, which is constrained to $b_n$.  
        When fitting $\pi_{b_n},$ which is a prediction model, we recommend that one follows established guidelines \cite{steyerberg2010assessing,collins2015transparent} for  estimating  and reporting prediction models. This might include assessment of the {behavioral policy model}  on held out data, as we have included in Figure \ref{calcurve}.  Note that a calibration curve (computed on held out data) helps us assess the reasonableness of the model specification and the estimation;  there is an associated calibration curve shape for a model that is too simple for the true data generating mechanism (an s-shaped curve) and an associated calibration curve shape for an estimation procedure that is overfitting (an s-shaped curve that is reflected over the identity line). \cite{van2016calibration,harrell2001regression,niculescu2005predicting}  In our case, since we assume that the behavioral policy is linear in the parameters by Equation (\ref{eq:bpolicy}), we  plot a calibration curve to assess whether linearity is too restrictive.   In this study, we also reduced noise in estimation of $b_n$ by using a penalized estimator,\cite{hastie2009elements}   where the penalty was chosen by cross validation (this can be done with the CV.GLMNET\cite{Friedman2010-rt} R package).  
        %We further resampled the real data multiple times and took an average, and this average was less noisy than each of its individual components.  %Finally, we assessed our estimate of $b_n$ by plotting calibration curves that were computed on test data within each resampling, and these calibration curves were also averaged. 
    
    }

\section{Simulations}
\label{rs:sim}
\subsection{Scenario}
In our simulations, we will investigate a problem inspired by  inpatient blood pressure control. In particular, let us consider inpatient hypotension management, which will be the focus of our real data analysis in Section \ref{rs:realdata}.   Suppose, as will be the case in the real data analysis, we have determined that healthcare providers take into account $K=9$ covariates when making a decision.  Suppose that we have also determined that hypotension management is a high-stakes decision problem, {so we fix $V^{min}$ to be the value that is 2 standard errors above the standard of care, and we suppose that} the healthcare providers will only adopt a policy that diverges from the standard of care for approximately one covariate.  We split the data into a training and test set. We use the training set to better understand how many coefficients will diverge for each $\lambda$, allowing us to assess the feasibility of our requirement on {value, $V^{min}$, and} the number of diverging coefficients, $C$, and we use the test set to obtain a valid estimate of value, $V_0$.

Fix the sample size to be ${n=1000},$ the number of Monte-Carlo repetitions to be $M=500,$ and the states to be $S_0,S_1\in \mathbb{R}^9$.  Recall from Equation (\ref{eq:thetapolicy}) that \[\pi_{b_0}(A_0=1|S_0=s_0)=\expit(b_0^Ts_0),\] and set the parameter of the true behavioral policy  to be {$b_0=(-0.01,0.02,0,\dots,0)^T$.}  Hence, although we have {\it a priori} suspected that 9 covariates are relevant to the decision problem, the true behavioral actors  base their decisions only on the first two covariates.

\subsection{Data generation}
\label{sec:dataGen}
Draw the initial state $\ S_0\sim \text{N}(\mu_0,\Sigma_0)$ and draw the action from a Bernoulli distribution, \[A_0|S_0 \sim \text{Bern}\left(\pi_{b_0}(A_0=1|S_0=s_0)\right),\] where {$\mu_0=(45,\dots,45)^T$} and $\Sigma_0,$ the initial state covariance, is a $K$-dimensional  matrix with {300 on its diagonals} except for covariance  of {100} between $S_{0,1}$ and $S_{0,2}$. {We choose this initial state to make the problem directly interpretable in terms of mean arterial  pressure (MAP), as in the real data analysis.}  Draw the final state 
\begin{equation}
\label{eq:transitionSim}
S_{1}|A_0,S_0\sim \text{N}(S_0 + \tau {^T} S_0 A_0 ,\Sigma ),
\end{equation}
where $\Sigma,$ the transition covariance, is a $K$-dimensional matrix with  {300 on its diagonals}  except for covariance of  {200}  between $S_{1,1}$ and $S_{1,2}.$ Let the coefficient for the action in Equation (\ref{eq:transitionSim}) be {$\tau=(0.1,0.7,0,\dots,0),$} so that only the {first and} second covariate of the state affect the transition.  Note that $\tau$ is also called the  ``treatment effect.''
%For example, we have in this case that a high value of $(S_t)_2$ leads to a larger increase in blood pressure; this might be the case if for example vasopressors exert the majority of their effect on working vessels, and therefore will have a larger treatment effect when blood pressure is higher.  When blood pressure is extremely low, this may indicate a deficiency in function of the vessels due to, for example, immune infiltrates, and administering a vasopressor might not easily increase blood pressure in this setting.  This is an oversimplification, and it is possible that in real data we would see the opposite (that vasopressors have greater effect when blood pressure is low), but 
 Set the reward function to be 
 \begin{equation}
 \label{eq:simReward}
      R(S_0,A_0,S_{1})=S_{1,2}. 
 \end{equation}
     {
        Hence, the transition will depend on two covariates (and hence the reward, which depends on the transition because the transition leads to the final state, will depend on two covariates). } 
 If we imagine that the second covariate is blood pressure, then this reward reflects our goal of  raising blood pressure, {where blood pressure might depend on both the past blood pressure and another covariate, such as heart rate}.  {Since we have made the states in the range of a typical MAP, we will now be able interpret the expected reward as the expected MAP.}
 
 The following fact will guide us in ensuring that our simulation results are reasonable. For the reward in our simulations, if the true optimizer of the unpenalized objective is $\beta_0=\arg\max_{\beta}V_0,$ as in Equation (\ref{eq:v0}), then, {for $\zeta_1$ and/or $\zeta_2$  arbitrarily large,}
\begin{equation}
\label{eq:sim.application}
\beta_{0}=({\zeta_1,\zeta_2},0,\dots,0).
\end{equation} 
Equation (\ref{eq:sim.application}), which is derived in Appendix \ref{app:proofSimBetaInf}, follows from the definition of the reward in Equation (\ref{eq:simReward}), an application of Equation (\ref{eq:determinism}), and our choice of treatment effect $\tau$.  
\newcommand{\simulationBetaInf}{
%\begin{proof}
Apply Equation (\ref{eq:determinism})  to obtain

\begin{align*}
\pi_0\left(A_0=1 \mid s_0\right) & =I\left(\left(E\left(R \mid S_0=s_0, A_0=1\right)-E\left(R \mid S_0=s_0, A_0=0\right)\right)>0\right) \\
& =I\left(\left(E\left(S_1 \mid S_0=s_0, A_0=1\right)-E\left(S_1 \mid S_0=s_0, A_0=0\right)\right)>0\right) \\
& =I\left(\left(S_0+\tau^{T} S_0 A_0-S_0\right)>0\right) \\
& =I\left({\tau_1 S_{0,1}}+{\tau_2} S_{0,2}>0\right) .
\end{align*}

We must now take a policy of the form of Equation (\ref{eq:bpolicy}), \expit( $\left.\beta_0^T s_0\right)$, and make it as close as possible to $I\left({\tau_1 S_{0,1}+\tau_2} S_{0,2}>0\right)$. We have that
\[
\expit\left(\beta_0^T s_0\right)=I\left({\tau_1 S_{0,1}+\tau_2} S_{0,2}>0\right) \iff {\beta_{0,1}} \text { and/or } \beta_{0,2} \text { are arbitrarily large. }
\]
Note that {we do not} have {$\beta_{0,1}=\infty$ or $\beta_{0,2}=\infty$, but instead just that $\beta_{0,1}$ or $\beta_{0,2}$ are arbitrarily large, since otherwise we encounter issues with linear predictors being undefined,} as discussed at the end of Appendix \ref{app:deterImpInf}. {In practice, our estimate $\beta_{n, 2}$ should dominate $\beta_{n, 1}$ in magnitude, since $\tau_2>>\tau_1$, and hence $S_{0,2}$ contributes more to the reward. We therefore expect $\beta_{n, 2}$ to be selected more often than $\beta_{n, 1}$, although $\beta_{n, 1}$ should be selected more often than if $\tau_1=0$.}
%\end{proof}
}
Hence, since {$\beta_{0,1}$and} $\beta_{0,2}$ {tend toward infinity}, we expect that {$\beta_{0,\lambda,1}$ and} $\beta_{0,\lambda,2}$ will have positive signs (note also that $b_{0,2}>0$, so even when $\lambda$ is large, $\beta_{0,\lambda,2}>0$, {and the opposite for $b_{0,1}$ and $\beta_{0,\lambda,1}$}), but,  by Lemma \ref{lemma:finiteBeta0lambda}, {since $L_1$ shrinks, neither} will be {arbitrarily large}.

\newcommand{\IPTW}{
\[E(R(1))-E(R(0))\approx \frac{1}{n}\sum_{i=1}^n \frac{R_i A_i}{\pi_{b_n}(A_i|S_i)}-\frac{1}{n}\sum_{i=1}^n \frac{R_i (1-A_i)}{\pi_{b_n}(A_i|S_i)}\]
}
\newcommand{\simpostable}{

\begin{table}[htp!]
\centering
\begin{tabular}{rl}
  \hline
 &  \\ 
  \hline
n & 1000 \\ 
  T & 2 \\ 
  %Prop. R<0 & 0.3 \\ 
  Prop A=1 & 0.489 \\ 
  min $\pi_b$ & 0.228 \\ 
  mean $\pi_b$ & 0.495 \\ 
  max $\pi_b$ & 0.761 \\ 
  IPTW treat. eff. & -0.3 \\ 
 % Par. treat. eff (p-val) & -0.153 (0) \\ 
   \hline
   \caption{Simulated data summary statistics. }
\label{simdatasum}
\end{tabular}
\end{table}
}

\newcommand{\notzerospiel}{{Note that we do not start the $\lambda$ grid at 0, since the unconstrained version is difficult to optimize, and the coefficient magnitudes for the unconstrained version are very large, so  the  range of the y-axis becomes large as well, making it difficult to visualize the changes in coefficients within the desired $\lambda$ range.}}
\subsection{Results}
\label{sec:sim.res}
\begin{figure}[htp!]
\centering
\includegraphics[width=0.70\textwidth]{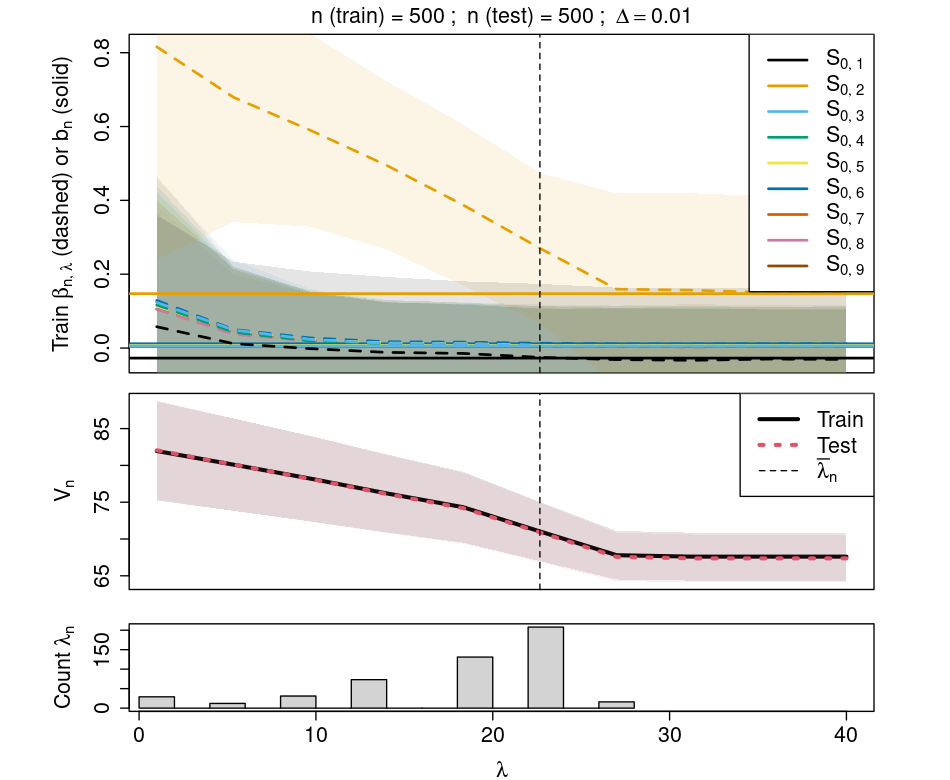}
	\caption{
{\bf Average coefficients ($\beta_{n,\lambda}, b_n$) and value ($V_n$) in the 500 simulated data sets.}
 In the top panel, the solid, horizontal lines correspond to the coefficients, $b_n$, of the behavioral policy, $\pi_{b_n}$, and the changing, dotted lines correspond to the coefficients, $\beta_{n,\lambda}$, of the suggested policy, $\pi_{\beta_{n,\lambda}}$.  The vertical dotted line, $\bar{\lambda}_n$, indicates the choice of $\lambda$ based on Equation (\ref{eq:lambda.crit.diff}) when using the average of the coefficients over Monte-Carlo datasets. The bottom panel shows the distribution of the different $\lambda_n$ that are selected using Equation (\ref{eq:lambda.crit.diff})  over the Monte-Carlo datasets. Note that $\Delta$ is the threshold for considering a coefficient equal to its behavioral counterpart. \notzerospiel}
\label{fig.sel.mc}
\end{figure}
We see in Figure \ref{fig.sel.mc}, which shows the average coefficients over 500 datasets, that, as $\lambda$ increases along the horizontal axis, the relative sparsity penalty  shrinks reward-irrelevant coefficients to their behavioral counterparts. We show confidence intervals for the coefficients $\beta_{n,\lambda}$ using a Monte-Carlo standard error estimate, as described in Appendix \ref{mcCIbetanlambda}.
  For the confidence intervals around $V_n$, we use the estimator described in Appendix \ref{sigmaV} and take an average of these confidence intervals over datasets.
  %In Figure \ref{fig.sel.mc}, we show how our criterion in Equation (\ref{eq:lambda.crit.diff}) can help us to select $\lambda$. 
  The coefficients in the top panel are the averages over datasets, and the $\bar{\lambda}_n$ selection in the middle panel of Figure \ref{fig.sel.mc} is made by optimizing the criterion in Equation (\ref{eq:lambda.crit.diff}) for these average coefficients (we do this once with the averages, to show how one selection would appear based on the top panel, and we also do this for each dataset, to show the distribution of selections).  The $\bar{\lambda}_n$ selection on the average coefficients occurs when one coefficient is non-behavioral, since $C=1.$  Since this is a simulation, and we generate multiple datasets, we also employ our $\lambda$ selection criterion in Equation (\ref{eq:lambda.crit.diff})  for each of the individual Monte-Carlo datasets, giving $M$ selections of $\lambda_n$, and we show the distribution of $\lambda_n$ in the bottom panel of Figure \ref{fig.sel.mc}.  Based on the $\lambda_n$ that is selected in each dataset, we have corresponding covariates that are selected (i.e., their coefficients are not set to their behavioral counterparts).  

 {We show the distribution over simulated datasets of the value function $V_n,$ the mean estimated probability of treatment (and its closeness to behavior), and the degree of relative sparsity.  We show this for the true optimal policy, the unconstrained optimization of $V_n$,   the constrained optimization of $J_n$ (with selected $\lambda_n$), and the behavioral policy, in Figure \ref{fig:comp.dists}.  We see that, as expected, the value without a penalty is higher on average, because we lose some value by constraining toward behavior, but we gain proximity to behavior and relative sparsity in doing so. Note still that, { over Monte-Carlo datasets, the distribution of which is shown in Figure \ref{fig:comp.dists}}, the observed/ behavioral policy has expected blood pressure of  {$V_n=67.60$} and that the suggested policy has expected blood pressure  of {$V_n=74.12$}, where the latter is at least two standard errors above the observed policy.  Note that other methods that find an optimal policy, but do not require modeling of the behavioral policy, would give the same solution as the top row of Figure \ref{fig:comp.dists}, and they do not allow us to obtain relative sparsity or closeness to behavior, since neither the behavioral nor the suggested policy is  modeled directly, and any closeness to behavior precludes optimality. Methods like TRPO (Equation \ref{eq:trust.pol.obj.sch}) allow us to obtain closeness to behavior, but not relative sparsity, since the KL divergence penalty does not impose sparsity (for more discussion, see Table \ref{qual}).}
\newcommand{\distspiel}{for the suggested policy ($\beta_{n,\lambda_n},$ where $\beta_{n,\lambda}=\arg\max_{\beta}J_n$); and for the behavioral policy, $b_n=\arg\max l_n$, where $V_n,J_n,$ and $l_n$ are value, penalized value, and log likelihood, respectively. 
Note that the suggested policy provides value, closeness to behavior, and relative sparsity, all of which facilitate adoption of the suggested policy into the clinic.}
  \begin{figure}
      \centering
      \includegraphics[width=0.9\textwidth]{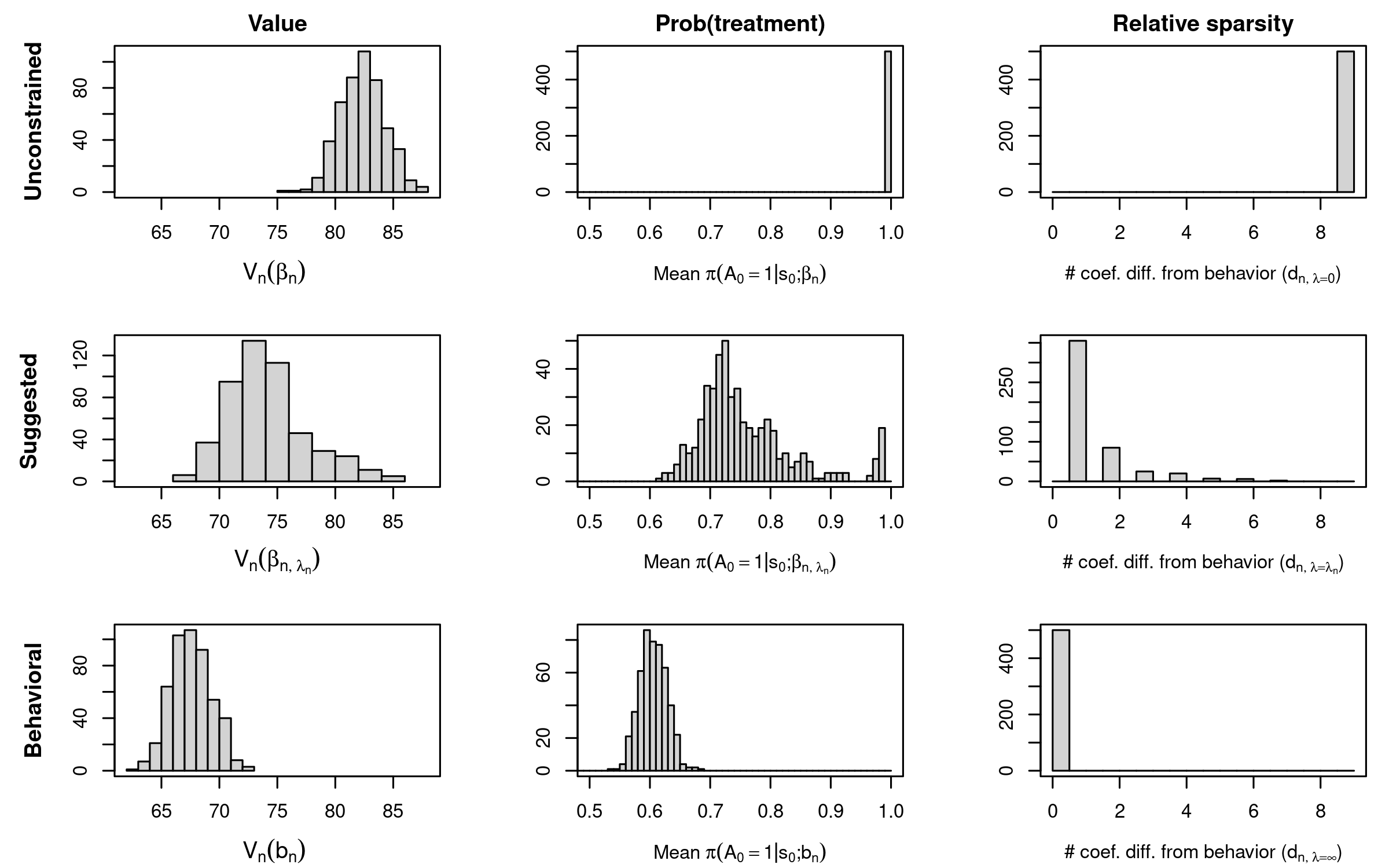}
      \caption{ {\bf {Empirical distributions} of value, probability of treatment, and relative sparsity for data policies over simulated datasets.}
      {We show the distribution over simulated datasets of the value, mean estimated probability of assigning treatment (and its proximity to behavior), and relative sparsity.  We show these plots %for the true maximizer $\beta_0=\arg\max_{\beta}V_0$, which we know from Equation (\ref{eq:sim.application}) (this would be the result of any method that just finds the optimal policy, and allows neither closeness to behavior nor relative sparsity); 
      for  unconstrained optimization ($\beta_n=\arg\max_{\beta} V_n$
      %which has distributions that are identical to the true maximizer, $\beta_0=\arg\max_{\beta}V_0$, which we know from Equation (\ref{eq:sim.application}) 
      , which would be the result of any reinforcement learning method that just finds the optimal policy, and allows neither closeness to behavior nor relative sparsity); \distspiel} %{Relative sparsity is omitted for the first row, because the true optimizer is a nonparametric indicator function (see Section \ref{app:indicator}), and therefore relative sparsity cannot be defined with respect to the parametric behavioral policy.}
      }, 
      \label{fig:comp.dists}
  \end{figure}
% latex table generated in R 4.2.1 by xtable 1.8-4 package
% Mon Jan 16 03:34:29 2023
\begin{table}[ht]
\centering
\begin{tabular}{rrrrrrrrrr}
  \hline
 & $S_{0,1}$ & $S_{0,2}$ & $S_{0,3}$ & $S_{0,4}$ & $S_{0,5}$ & $S_{0,6}$ & $S_{0,7}$ & $S_{0,8}$ & $S_{0,9}$ \\ 
  \hline
$\hat{P}(selected)$ & 0.27 & 0.97 & 0.12 & 0.10 & 0.12 & 0.10 & 0.10 & 0.10 & 0.11 \\ 
   \hline
\end{tabular}
\caption{{\bf Selection proportion for each covariate over 500 simulated datasets.}  Note that only $S_{0,2}$ {and, to a lesser extent, $S_{0,1},$} are truly relevant to the reward, and that there is some correlation between $S_{0,1}$ and $S_{0,2}$.}
\label{tb:sel}
\end{table}
 {To better characterize the relative sparsity that we observe in the policies estimated on the simulated datasets, we} include the distribution of $D_{n,\lambda_n},$ the number of  selected coefficients {over} Monte-Carlo datasets, as defined by Equation (\ref{eq:lambda.crit.diff}), in {Figure \ref{fig:comp.dists}}. We see that  $D_{n,\lambda_n}$, the number of diverging parameters under the selected $\lambda_n,$  matches our target of $C=1$ more than half of the time, and that we approximately achieve our goal of $C=1$ in all cases.  {Sometimes we also select $D_{n,\lambda_n}=2,$ due to the correlation between $S_{0,1}$ and $S_{0,2}.$}   We report the covariate selection proportions in Table \ref{tb:sel}, where we see that the {most} reward-relevant covariate, $S_{0,2},$ is most often selected  {(along with sometimes the covariate $S_{0,1}$, which is correlated with $S_{0,2}$ and more faintly related to the reward. The other covariates are selected less often}).

%   \begin{table}[ht]
% \centering
% \begin{tabular}{rlllllll}
%   \hline
% $d_{n,\lambda_n}$ & 1 & 2 & 3 & 4 & 5 & 6 & 8 \\ 
%   \hline
%   $\hat{P}(D_{n,\lambda_n}=d_{n,\lambda_n})$ & 0.446 & 0.238 & 0.142 & 0.090 & 0.048 & 0.034 & 0.002 \\ 
%    \hline
% \end{tabular}
% \caption{ {\bf Observed distribution of $D_{\lambda_n},$ the number coefficients that diverge from their behavioral counterparts for a selected $\lambda_n$,
%                over 500 Monte-Carlo datasets.} }
% \label{tab:Cn}
% \end{table}

\section{Real data analysis }
\label{rs:realdata}
\subsection{Research objective}
\label{real.data.problem.statement}
There is variability in vasopressor administration for hypotensive patients in the intensive care unit (ICU), and trials on vasopressors have been  inconclusive. \citep{lee2012interrogating,der2020narrative,russell2021vasopressor}  Vasopressor usage is therefore an interesting potential target for medical decision models.   Vasopressor use in the inpatient setting can impact a patient greatly, and the onset of action is often in minutes. Vasopressors can  stabilize blood pressure, but they have a variety of adverse effects. Secondary to excessive vasoconstriction, vasopressors can cause organ ischemia and infarction, hyperglycemia, hyperlactatemia, tachycardia, and tachyarrhythmias. \citep{russell2021vasopressor}     Whether to prescribe vasopressors is a subtle and high stakes decision.   To change behavior with respect to vasopressor usage, therefore, any divergence from the established care guidelines should be clear and justifiable.  Let us therefore try to obtain a policy {that has value at least 2 standard errors above the behavioral policy, but} that diverges from behavior with respect to only 1 or 2 covariates.  
%Using the relative sparsity objective in Equation (\ref{eq:jn}) and the criterion for selecting $\lambda$ in Equation (\ref{eq:lambda.crit.diff}), we will show how we can obtain a policy that has higher value than the behavioral policy but that diverges from the established care policy only with respect to the coefficients for roughly 2 covariates, facilitating the explanation of the policy and helping to justify its implementation.
%\subsection{Dataset}
\subsection{Dataset and cohort selection}
We consider the MIMIC III dataset,  \citep{johnson2016mimic2, johnson2016mimic1,goldberger2000physiobank} which is a freely available, observational electronic health record dataset from the  Beth Israel Deaconess Medical Center. We briefly describe cohort selection. We consider only the medical ICU (i.e., we do not consider the surgical or trauma ICUs).  If a patient has been hospitalized multiple times, we take the first hospitalization, and if a patient is admitted to the medical ICU multiple times within one hospital stay, we take the first medical ICU stay. In our decision problem, which we will describe in detail in Section \ref{sect:realDataDecisionProb}, we will analyze a time window that begins at hypotension onset and lasts 30 minutes. We excluded the 7 out of $n=4,715$ patients who left the ICU before those 30 minutes had elapsed.
%We consider only the medical ICU (we do not add patients from, e.g., the cardiac and surgical ICUs, since we consider those patients to be generally different from the medical ICU patients).  

\subsection{Decision problem}
\label{sect:realDataDecisionProb}
We amend code from Futoma et al. \citep{futoma2020popcorn} to obtain a decision problem that begins approximately at the onset of hypotension. Hypotension is assessed according to mean arterial pressure (MAP), a weighted average of diastolic and systolic blood pressures, where the weights reflect the amount of time in diastole and systole; MAP is also the product of cardiac output and total peripheral vascular resistance.  We define time zero to be approximately  at the first MAP$<60$, which is a threshold for  hypotension. \citep{yapps2017hypotension,lee2012interrogating}  After 15 minutes, we construct $S_0,$ which contains a summary of all of the covariates at this time point.  We found it was better to let $S_0$ be a summary of the first 15 minutes instead of the observed MAP$<$60 itself, because one MAP$<60$ may be influenced by random fluctuations, whereas if MAP$<60$ at time zero and is still $<60$ after 15 minutes, then it is likely that the patient is experiencing a sustained hypotensive episode.  We use the set of covariates  from Futoma et al., \citep{futoma2020popcorn} which includes MAP, heart rate (HR), urine output, lactate, Glasgow coma score (GCS), serum creatinine, fraction of inspired oxygen (FiO2), total bilirubin, and platelet count.  These  covariates would all be of interest when deciding whether to administer vasopressors. For MAP, if there was more than one measurement within a time interval,  we used the most recent, assuming the most recent  measurement would be most relevant as a reward and when deciding whether to administer vasopressors in the next time step. These covariates, taken as a vector, define the state. The action $A_0,$ based on $S_0$, is  whether to administer vasopressors from time 15 minutes to time 30 minutes.  The final state, $S_1,$ contains a summary of  MAP and all other covariates at 30 minutes.

The vasopressor units are total micrograms of medication given each hour per kilogram of body weight. In particular, as in Futoma et al., \citep{futoma2020popcorn} we consider Dopamine, Epinephrine, Norepinephrine, Vasopressin, and Phenylephrine.
The various doses and brands of vasopressors are converted based on the code in Futoma et al. \citep{futoma2020popcorn} to a Norepinephrine equivalent using the method in Komorowski et al. \citep{komorowski2018artificial}  Intravenous Norepinephrine has a half life of approximately 2.4 minutes. \citep{smith2021norepinephrine}   Hence, we consider the MAP at 30 minutes to depend only on the vasopressors administered from 15-30 minutes, and not on those administered from 0-15 minutes.  
%Hence, our policy depends only on the state at 15 minutes, and not on the action taken from 0-15 minutes (also, this action is somewhat summarized in the MAP at 15 minutes, which is present in $S_0$).  
Since we analyze intensive-care-unit data, most patients are started on vasopressors before the 15 minute time point, in which case our decision becomes whether to continue or to stop the medication.
We finally define a reward that reflects the short term goal of increasing MAP in the setting of severe hypotension. In our real data analysis, we took MAP to be the first covariate in the state, so we  define the reward to be $R(S_0,A_0,S_1)=S_{1,1}.$ Vasopressors should increase this reward. 

As an example of a trajectory, one patient was found to have MAP=58, which indicated hypotension. Vasopressor infusion was  initiated within minutes.  From 0-15 minutes, as the patient received the vasopressor infusion, their MAP changed from 58 to 48.  Hence, since we take MAP to be the first covariate, we had MAP=48 as $S_{0,1}$. From 15-30 minutes,  the vasopressor infusion was maintained, so $A_0=1$.   Finally, at 30 minutes,  MAP=53, so $S_{1,1}=53$.   Hence, this patient had trajectory $(S_{0,1}=48, A_0=1,S_{1,1}=53)$ and $R(S_0,A_0,S_1)=53$ (in practice, the state covariates are scaled, as described in Section \ref{section:relativeSparsity} and Appendix \ref{scaling}, but this is essentially the trajectory).

% and normalize each as in \citet{komorowski2018artificial} so that  different types and doses of vasopressors are converted to comparable units.   

  %{Note that if we define the action taken before $A_0$ to be $A_{-1},$ then we do not adjust for $A_{-1}$ in the behavioral policy as a confounder, because, by the standard MDP assumptions, the transition probability $S_1|A_0,S_0$ depends only on $A_0,S_0.$ In other words, the standard MDP assumptions consider $S_0$ a sufficient summary of $A_{-1}.$}
The measurements that contribute to the covariates are provided in extensive tables that are part of the MIMIC-III database (e.g., one of the tables has roughly 330 million rows), which were processed  using bash scripts from Futoma et al. \citep{futoma2020popcorn} As in Futoma et al., \citep{futoma2020popcorn} extreme, non-physiologically feasible values of covariates were floored or capped, missing static covariates were imputed by the median, and time dependent covariates were imputed by last observation carry forward.

\subsection{Results  }
\label{sec:realdataresults}
%We show some relevant characteristics of the real data in Table \ref{realdatadesc}.  
%We checked that $\min_s\pi_b(A=a|s)>0, \forall a,$ which implies that Assumption \ref{assum:pos} is empirically valid. We also checked that we have a strong
%positive treatment effect that corresponds to the known physiological effect of vasopressors on blood pressure  ({Todo, get p-val for IPTW (not shown, parametric is positive and sig).})

% latex table generated in R 4.0.2 by xtable 1.8-4 package
%% Fri Jun 24 17:27:48 2022
%\begin{table}[htp!]
%\centering
%\begin{tabular}{rl}
%  \hline
% &  \\ 
%  \hline
%n & 4708 \\ 
%  T & 2 \\ 
  %Prop. R<0 & 0.043 \\ 
%   Proportion with A=1 & 0.146 \\ 
%   Min $\pi_b(a|s)$ & 0.016 \\ 
%   Mean $\pi_b(a|s)$ & 0.153 \\ 
%   Max $\pi_b(a|s)$ & 0.971 \\ 
%   IPTW treat. eff. & 0.085 \\ 
%   %Par. treat. eff (p-val) & 0.044 (0) \\ 
%   \hline
% \end{tabular}
% \caption{Characteristics of the real data.  $T=$ number of stages, $\pi_b$ is the behavioral policy of the providers, $IPTW$ is inverse probability weighted treatment effect (see Appendix were taken over the behavioral policy treatment probabilities for all $nT$ time steps in the data.}
% %, and Par treatment effect is the coefficient and p-value on $A$ in a linear model adjusting for counfounders and the action
% .
%  }
% \label{realdatadesc}
% \end{table}

% We show box plots representing the marginal treatment effect for each stage in Figure \ref{real.data.stagerewards}.  We can visualize that  the majority of the increase in reward is present in the first roughly 15 minutes after hypotension determination.

\begin{figure}[htp!]
\centering
% this is in sam_slides_docs/7-24-22_WithMimic/mimic
\includegraphics[width=0.70\textwidth]{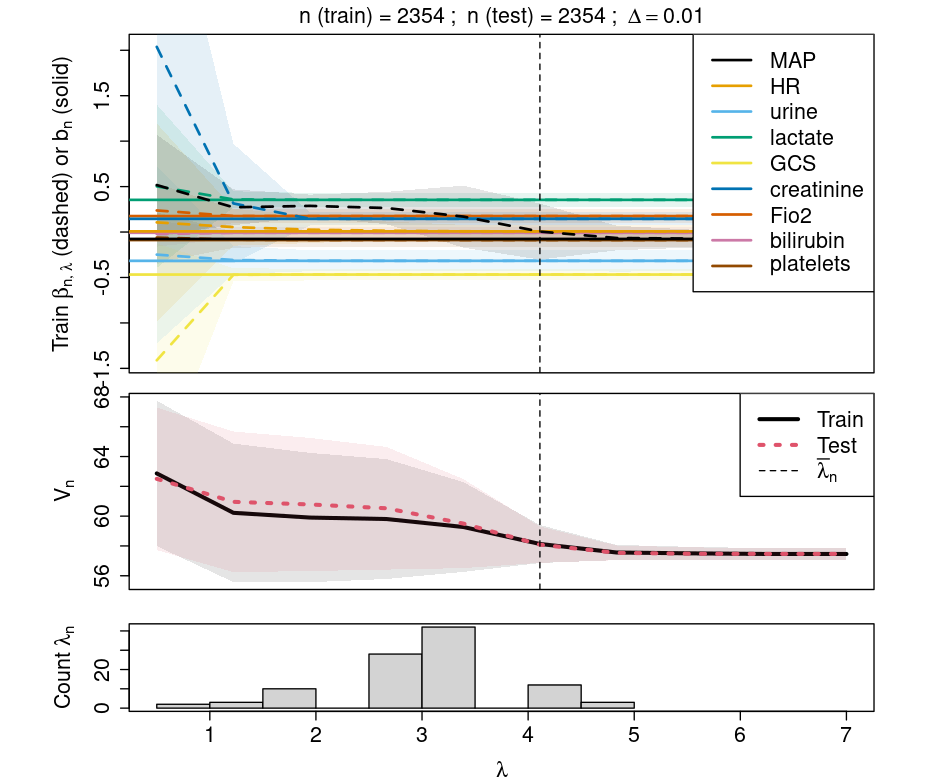}
\caption{{\bf  Coefficients ($\beta_{n,\lambda}, b_n$) and value ($V_n$) in the real data set.}  In the top panel, the solid, horizontal lines correspond to the coefficients, $b_n$, of the behavioral policy, $\pi_{b_n}$, and the changing, dotted lines correspond to the coefficients, $\beta_{n,\lambda}$, of the suggested policy, $\pi_{\beta_{n,\lambda}}$.   The dotted vertical line in the bottom panel indicates a choice of $\lambda$ based on maximizing Equation (\ref{eq:lambda.crit.diff}) in the training data.  Note that $\Delta$ is the threshold for considering a coefficient equal to behavior  and was chosen based on visual inspection of the top panel. {We average over 100  resamples of the real data; we show the distribution of the selected $\lambda_n$ over resamples, but the final selection of the penalty weight, $\bar{\lambda}_n$, was based on the average coefficients and value over resamples.} \notzerospiel {Note that these confidence intervals for the coefficients, which serve just as a visualization of uncertainty, have different meaning under Monte-Carlo resampling than under repeated sampling of the real data (for more details on the  resampling-based construction of confidence intervals for the coefficients, see Appendix \ref{mcCIbetanlambda}).} }
\label{real.data.lam.grid}
\end{figure}

We first evaluate the specification of the model for the behavioral policy in Equation (\ref{eq:bpolicy}), which is integral to inverse probability weighting in Equation (\ref{eq:vnest}) and also to the penalty in Equation (\ref{eq:jn}). For this, we show a calibration curve (Figure \ref{calcurve} of Section \ref{app:calc}). We see high concordance between the observed and estimated probabilities, and therefore we conclude that our model for the behavioral policy, Equation  (\ref{eq:bpolicy}), is a reasonable model, and that we can then proceed with optimizing Equation (\ref{eq:jn}) for the proposed method. Since we only have one real dataset, unlike in simulations, we split the data into single train and test sets, but we repeat this split 100 times and average the results. We recommend that this resampling be done if the objective function is not otherwise\cite{thomas2015safe} stabilized. 
We show the coefficient paths as a function of $\lambda$ in Figure \ref{real.data.lam.grid}. We see that the coefficient for { MAP }withstands the relative sparsity penalty, requiring a large $\lambda$ to finally reach its behavioral counterpart, whereas coefficients for variables like platelet count quickly approach their behavioral counterparts, causing virtually no change in value. {To best assess which coefficients withstand the push to behavior, and how this impacts the value, $V_n$,  it is vital to perform some type of repeated sampling, as we have done here, because the individual test train splits can be noisy. One can also stabilize $V_n$ in other ways,\cite{thomas2015safe} which might be less computationally intensive.}
%We see that some variables hold out against the relative sparsity penalty, approaching their behavioral counterparts slowly, and that others rapidly approach their behavioral counterparts.  We see that when some variables approach their behavioral counterparts, value changes greatly, and that when other variables approach their behavioral counterparts, value is roughly constant.  We see that our selection criterion in Equation (\ref{eq:lambda.crit.diff}) gives us a policy that diverges from behavior with respect to roughly 2 covariates (actually 1 covariate), as prespecified, but also gives the maximal value of the policies that diverge with respect to 2 covariates.  This gives a policy that achieves both relative sparsity and an increase in value with respect to the behavioral policy.  In particular, w

  \begin{figure}
      \centering
      \includegraphics[width=0.9\textwidth]{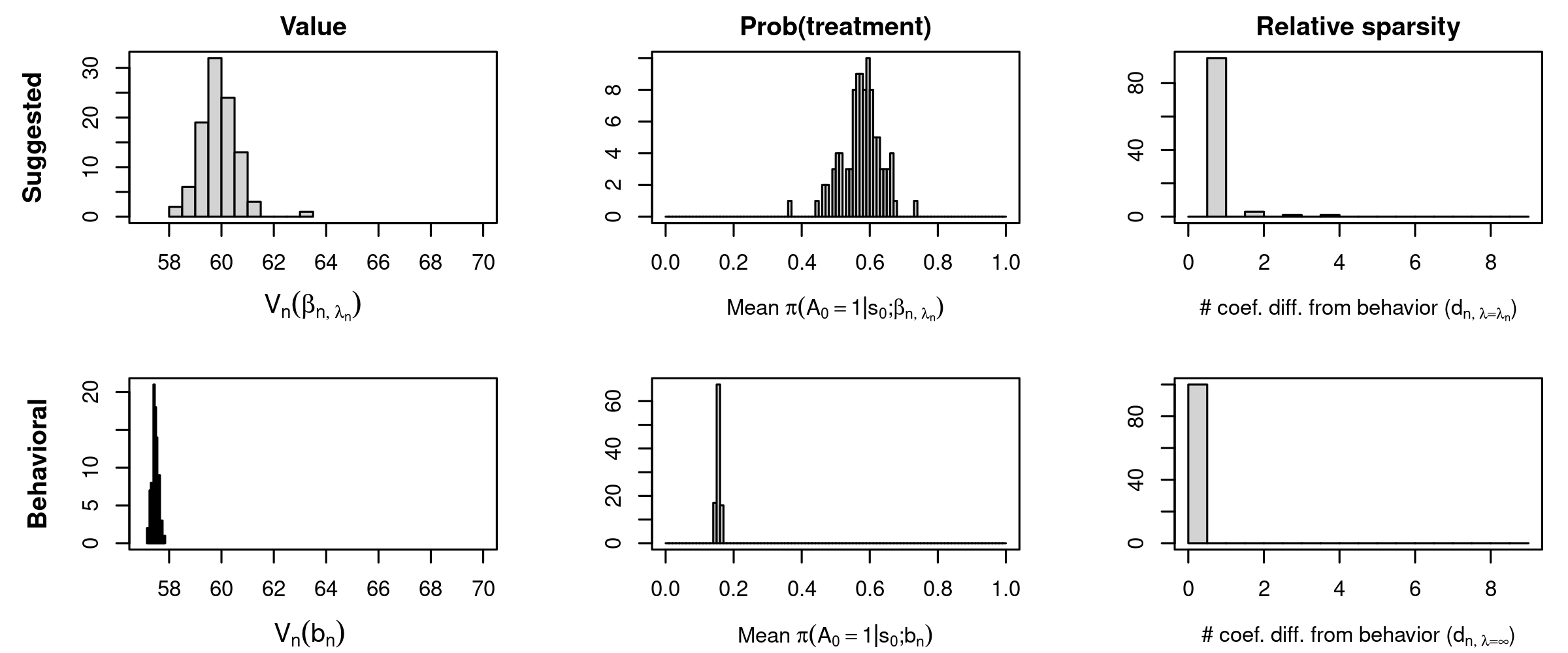}
      \caption{ {\bf  {Empirical distributions} of value, mean probability of treatment, and relative sparsity for data policies over resamples of the real data.}
      {We show the distribution over resampled datasets of the value, mean estimated probability of assigning treatment (and its proximity to behavior), and relative sparsity.  We show these plots \distspiel}}
      \label{fig:comp.dists.real}
  \end{figure}

Here, we determined that for the criterion in Equation (\ref{eq:lambda.crit.diff}), {given value, $V_n$, that is at least as great as $V^{min}$, $C=1$ diverging covariates would be acceptable.  Note that $\Delta$ is the threshold for considering a coefficient equal to behavior and was chosen based on visual inspection of the top panel. }
%(in practice, compared to the selected covariates, the contribution of these nearly non-selected covaraites to the linear predictor is similar to what their contribution would be if they were set to their behavioral values exactly).}
%Recall also that one might have specified $C=1$ and then viewed the top panel in Figure \ref{real.data.lam.grid} only to conclude that it is not possible to obtain a reasonable policy with $C=1$,  {because it is not possible to obtain $V_n$ greater than $V^{min}$ when only allowing one covariate to diverge.  In this case, one could instead target $C=2$ and obtain a policy with greater value, at the cost of some sparsity.   Adjusting $C$ as we describe is  done somewhat  automatically as well, because the criterion in Equation (\ref{eq:lambda.crit.diff}) allows some slack with respect to matching $C$ exactly, but we recommend that one always inspect coefficient plots.  
We ultimately see that we are able to illustrate the behavior of the relative sparsity penalty with real data, and that we obtain a policy that is sparse relative to the behavioral policy while on average increasing value, $V_n$. {In the real data, { over resampled datasets, the distributions of which are shown in Figure \ref{fig:comp.dists.real}}, the average final MAP when following the standard of care is {57.46}, while the average final MAP when following the suggested policy is {59.89}, an increase of more than two standard errors (and 60 is the limit at which the vital organs are adequately perfused \cite{demers2021physiology}).}  

In particular, to justify this suggested policy to providers and patients, we only need to discuss {MAP,} because the parameters for all of the other covariates align with the existing standard of care. {Recall that every patient in the cohort was determined to be hypotensive, and that we observe them for 15 minutes to obtain their initial MAP (which is used to determine whether to treat), and then we observe them for another 15 minutes to obtain their final MAP (the reward). {Note that the behavioral coefficient on the initial MAP is negative.} In the behavioral policy, the providers prefer to treat patients with lower initial MAP more than patients with higher initial MAP; this makes intuitive sense, {since} blood pressure should be treated when it is low. {In the unconstrained, reward-maximizing policy (or the ``optimal'' policy), the coefficient on initial MAP is positive.
The optimal policy would require that all patients be treated, and that patients with higher initial MAP be treated more than patients with lower MAP (within the cohort of hypotensive patients). %Note that we adjust for initial MAP, so this is not a matter of confounding, and perhaps it makes some sense, since patients with higher baseline MAP might respond better to exogeneous vasopressors, since they are responding to their own endogenous vasopressors. }
%(possibly this is due to a stronger physiological response to vasopressors in these patients, or because a high MAP after a hypotensive episode might be due to random variation, and if left untreated, the MAP in the next 15 minutes may be worse than the MAP for patient's who randomly had a lower MAP in the first 15 minutes and received treatment).
The suggested policy, { with a coefficient for MAP that is near zero}, is somewhere in the middle; it tends to treat patients with higher initial MAP along with patients with lower initial MAP, and this increases the average reward. {More discussion on the MAP coefficient as it jointly relates to residual confounding, the specified reward, and the decision problem, is provided in the next section.} %However, the suggested policy  is not as drastic as the unconstrained policy.  The suggested policy is a good strategy if we acknowledge that it is good to increase average MAP beyond what is given by the current standard of care (the average MAP when following the standard of care is below the threshold for perfusion of vital organs), but that it is important to respect the standard of care, which likely takes into consideration other factors that are not easily incorporated into a reward function. 
%In this case, one can think of the proposed method as interpolating between the behavioral and the ``optimal'' (i.e., specified-reward-maximizing) policy.  One can equally think of the proposed method as interpolating between the behavioral reward, which we do not observe, but which we trust to some extent, and the specified reward, which is sensible but can be problematic when its maximization is pursued in an unconstrained fashion.  %We hence improve care in a small, but significant, step.
%, without creating drastic changes that may be reflections of the simplicity of our specified reward, by hedging our bets with respect to the specified reward against the behavioral reward, but without having to estimate the behavioral reward directly.
   %In a sense, we interpolate not only between policies, but between rewards. In this case, the specified reward is just the final MAP. The ``op The behavioral reward clearly takes into account more than just MAP. It might be more sensible to increase MAP more than do the behavioral actors, but not to increase it as much as possible.  
%Note that the preceding discussion, in which we pinpoint one covariate that diverges from behavior and go into depth regarding its role in the treatment decision, is facilitated by the output of the relative sparsity approach.
}
		 
\section{Discussion}

\subsection{Summary}
 We show how the relative sparsity penalty can be used to obtain a policy that is easy to explain in the context of the current standard of care, and therefore readily justifiable.
 %in other words, the relative sparsity penalty can allow us to derive a policy that is easy to explain in the context of the current standard of care, helping us justify it to the end-users.  
 We contrast our work with existing KL-based behavior constraints,\citep{schulman2015trust} which have been primarily focused on applications in robotics, and in which the difference between the suggested and behavioral policy is a black box.   Our methodology, which makes the difference between the suggested and behavioral policy sparse, and therefore more interpretable, has strong practical implications for the adoption (and interrogation) of data-driven decision aids in a healthcare setting.  
 
\subsection{Limitations and future work}
\label{subsec:limFuture}
We first discuss some real data analysis limitations.   We emphasize that any policy derived with our method  should be reviewed by the medical care team, and we  hope that  our method, in its interpretability, facilitates this type of review.

%We consider the early management of hypotension  (the first 30 minutes), because it is possible that in this times the goal to increase blood pressure outweighs other considerations, but ideally we would still take mortality and morbidity into account. 

 {We discuss the implications and plausibility of the assumptions needed for causal identification.  A major assumption is that we have measured all  confounders.  }
Our list of covariates ({given in Section \ref{sect:realDataDecisionProb})}, which match those in Futoma et al., \citep{futoma2020popcorn} appears to be quite comprehensive for this problem, but it is also important  to recognize that the no unmeasured confounders assumption cannot be tested empirically.
{
We note that the suggested policy has a positive, though near zero, coefficient on MAP, whereas the behavioral policy, or standard of care, has a negative coefficient on MAP.  %Note that the suggested policy is drawn toward the unconstrained reward-maxizing policy, which has a large positive coefficient on MAP. For the unconstrained reward-maximizing policy, given two hypotensive patients, the reward-maximizing decision is to prescribe to both, but to prescribe with higher probability to the patient with the baseline higher MAP. Prescribing with higher probabilty to the patient with lower MAP will lead to, on average, lower final MAP.
This may be because of residual confounding. There are different etiologies of hypotension (e.g., sepsis and hemorrhage \citep{standl2018nomenclature}), which are associated with different mechanisms by which the body becomes hypotensive (e.g., different effects on vasopressor-binding adrenergic receptors \citep{geevarghese2023role,de2009bench,standl2018nomenclature}), which might in turn alter the way the body responds to vasopressors.
%One can construct subgroups of patients  defined by their hypotension etiologies.
%Certain subgroups of patients might be less responsive to vasopressors, and  certain subgroups might have lower baseline MAP than other subgroups.  
%Compare the group of septic patients to those who are undergoing acute hypovolemic hypotension (e.g., due to hemorrhage), who, at least in the initial stages of shock, might have vasculature that behaves differently from the vasculature in septic patients, and which perhaps manifests as a tendency to respond more robustly to vasopressor therapy.  
It may be the case that one subgroup, e.g. with hypotension due to sepsis, responds less strongly to vasopressors and has a lower baseline MAP than another subgroup, e.g. with hypotension due to hemorrhage.
To maximize the average reward, or the average final blood pressure, over these two subgroups of patients, it makes sense to preferentially treat the  subgroup of patients who have stronger responses to vasopressors. 
%If the responsive patients have baseline higher MAP than other patients, we might see a positive coefficient on MAP in the reward-maximizing policy, to which the suggested policy is drawn. 
The standard of care, however, might  be more likely to give vasopressors to the subgroup experiencing sepsis than to the subgroup experiencing hemorrhage (generally, vasopressors and fluids are the first line treatment for sepsis, whereas there is more controversy about using vasopressors for hemorrhage \citep{gupta2017vasopressors,colling2018vasopressors}). In such a case, the signs on the MAP coefficient for the behavioral and reward-maximizing policy, to which the suggested policy is drawn, may  be different. Therefore, the subgroup to which patients belong may be an unmeasured confounder.  These subgroups might be quite complex and interact with patient characteristics.\citep{marin1995age} Although the covariates in Futoma et al.,\citep{futoma2020popcorn} which we use in our real data analysis, might fluctuate somewhat according to subgroup, a covariate that identifies these subgroups explicitly is not included. This may be because of the difficulty in establishing such subgroups. Sometimes, diagnoses such as ``sepsis'' are assigned at the end of the hospitalization (which makes it difficult to establish the time of onset), and such assignments, which are often generated for claims, can be of poor quality.\citep{rudrapatna2020accuracy}   It would be interesting to possibly use microbiology results to, e.g., determine sepsis status in future work.} The MIMIC dataset has many additional covariates, and considering the high-dimensional case might {also} be a direction of future work. 

{The specified reward, which is the final MAP in our case, might also be partially responsible for the positive coefficient for MAP in the reward-maximizing policy, to which the suggested policy is drawn.  In particular, the specified reward may be over simplified. Although it is sensible to increase MAP, a reward that is based only on increasing MAP may be overly simplistic because it leaves out other aspects that might also be important.} This specified reward could be improved by taking into account mortality and morbidity.\citep{lee2012interrogating}    { When taking into account mortality and morbidity, the reward-maximizing policy might be more discerning with respect to the coefficient on MAP.  A more indirect way to take into account mortality and morbidity using the proposed method would be to change the weight that we place on the specified reward by setting a smaller minimum value cutoff for the suggested policy, or a smaller $V^{min}$ in (\ref{eq:lambda.crit.diff}). By sacrificing some value, this allow us to choose a tuning parameter that yields a suggested policy that is closer to the standard of care. The standard of care optimizes its own reward, which likely takes into account mortality and morbidity, and, therefore, by choosing a policy that is close to the standard of care, we can do the same.  %Hence, the reward that we maximize with the suggested policy is somewhere in between the behavioral reward and the specified reward.

In general, the proposed method provides transparency about the coefficients that change when moving from the behavioral to the suggested policy. This generates useful discussion about the coefficients.  If the difference between a coefficient in the suggested policy and the standard of care is not clinically palpable, action must be taken before implementation of the policy into the clinic.  First, if it is suspected that confounders are unmeasured, these confounders should be collected if at all possible.  There will sometimes be evidence of this, because the standard of care coefficients might be unexpected, or, upon consult with a provider, we might find that they take into account some outcome-influencing variables that we had not collected. 
Second, if it is suspected that the reward function is overly simplistic, then one can discuss lowering the minimum value, $V^{min}$ in (\ref{eq:lambda.crit.diff}), in order to move toward the standard of care.  Third, one might also re-evaluate the decision problem.  In our case, for example, the problem is somewhat simplified; we have discretized time, and we have not taken into account mortality and morbidity in the reward. This has allowed us to avoid methodology needed for survival analysis\citep{cho2020multi} and continuous time, \citep{hua2022personalized} which could be the subject of future work. 
%We emphasize that the relative sparsity method spurs the three steps above, and enlivens the discussion, in ways that other more black-box methods may not.
}
Note {also} that reward function learning is an active area of research. \citep{ng2000algorithms,ramachandran2007bayesian}

        %We however have studied the major causes of hypotension, and we believe that the choices in \cite{futoma2020popcorn} encompass many of them; further, labs in the EHR are particularly granular and comprehensive (as opposed to e.g. administrative data, such as billing codes).  
       {The high concordance between observed and estimated probabilities seen in the calibration curve further supports the functional form posited for the behavioral policy in Equation (\ref{eq:bpolicy}), which we also consider to be a causal assumption, because the form of the behavioral policy plays such an important role in inverse probability weighting.}
       {For cases when Equation (\ref{eq:bpolicy}) does not hold, which would be detectable by a calibration curve, one could improve any issues with model misspecification by using basis expansions of the states.  This then allows us to fit a complex behavioral function that is still linear in the parameters, but not in the covariates {(e.g., consider $\beta_1 HR$ + $\beta_2 MAP$ vs. $\beta_1 HR$ + $\beta_2 HR^2 + \beta_3 MAP + \beta_4 MAP^2$)}. We could then map the bases to their respective covariates, and describe which covariates change from behavior accordingly. } Another direction for future work would be to consider decoupling, \citep{hilton2021batch} in which case we could make the behavioral policy in the value nonparametric and make the behavioral policy in the penalty parameteric (possibly with basis expansions).   
Such an addition might complicate convergence of the inverse probability weighted estimator, which is known to be sensitive to nonparametric nuisance parameters, especially in the high-dimensional case, but this may be acceptable with a large enough sample.   {This is a consequence of  slower convergence of the nonparametric estimator.\cite{ertefaie2020nonparametric}  Note that although there are other methods that allow one to sidestep the difficulty in specifying the behavioral policy altogether, these methods do not model the suggested policy directly, and therefore do not allow us to obtain relative sparsity between the coefficients of the behavioral policy and the suggested policy (for more discussion, see Section \ref{sect:l0l1} and Table \ref{qual}). {As in any modeling, overfitting is always a risk, for both the behavioral and suggested policy.  For the behavioral policy, we describe our method to prevent overfitting in Section \ref{behestim}. We believe that the overfitting issue for the suggested policy is mitigated by the parametric nature of the models and by the penalty toward behavior. 
      We believe that evidence for the suggested policy not overfitting is that the out of sample (test) value function (in Figures \ref{fig.sel.mc} and \ref{real.data.lam.grid}) corresponding to suggested policy is a nondecreasing function (and matches the training value function).  Specifically, if the suggested policy estimates were affected by overfitting, the resulting test value function might increase and then decrease (or be nonsmooth with several local minima and maxima). 
      }
}       %\calc

        {In terms of causal assumptions, we also assume positivity, or that in the behavioral policy/standard of care any action is possible given any state, which can be tested empirically, and we did so in the preliminary data analysis that accompanied this work.   Other causal assumptions include the stable unit treatment value assumption (SUTVA) and consistency of the states as potential outcomes of the actions, both of which are generally reasonable, since we observe the blood pressure under a certain treatment, and one patient's treatment does not influence another patient's blood pressure.   We have included some additional references\cite{kennedy2022semiparametric, munoz2012population,haneuse2013estimation,young2014identification,kennedy2019nonparametric} on causal identification in our setting. We emphasize that even with careful consideration of the plausibility of the causal assumptions, as we have attempted here, it is important  to conduct trials to evaluate any clinical decision aids before translation to routine care.
}

In conclusion, we present a method to obtain policies that are easy to explain in the context of the standard of care. We believe that using statistical methods to help solve decision problems in healthcare will be an iterative process involving collaboration among data analysts, healthcare providers, and patients.   We hope that the proposed methodology can help us better explain and justify data-driven policies to healthcare providers and patients, facilitating the adoption of these policies and invigorating the discussion. 
\section*{Acknowledgments}
The authors thank Joseph Futoma for providing code to preprocess and construct trajectories in the MIMIC data. The authors thank Jeremiah Jones, Ben Baer, Michael McDermott, Brent Johnson, Jesse Wang, Derick Peterson, and Kah Poh Loh for helpful discussions.  This research, which is the sole responsibility of the authors and not the National Institutes of Health (NIH), was supported by the National Institute of Environmental Health Sciences (NIEHS) and the National Institute of General Medical Sciences (NIGMS) under T32ES007271  and T32GM007356, respectively.

% for arxiv
\bibliographystyle{plainnat}
\bibliography{main}

% \subsection*{Author contributions}

% This is an author contribution text. This is an author contribution text. This is an author contribution text. This is an author contribution text. This is an author contribution text.

% \subsection*{Financial disclosure}

% None reported.

% \subsection*{Conflict of interest}

% The authors declare no potential conflict of interests.

% for arxiv
 %\bibliographystyle{plainnat}
 %\bibliography{main}

\section*{Supporting information}

Research code can be found at \url{https://github.com/samuelweisenthal/relative_sparsity}.  Code from Futoma et al. \citep{futoma2020popcorn} can be found at \url{https://github.com/dtak/POPCORN-POMDP}.
%can be found at
%\url{projects/Latent/sam/thesis_work/relative_sparsity/code_betareg/mdp_Vonly}.  
%The MIMIC data can be found at \url{https://physionet.org/content/mimiciii/1.4/}. 

\section*{Data availability statement}
The MIMIC \citep{johnson2016mimic} dataset that supports the findings of this study is openly available at PhysioNet (doi: \url{10.13026/C2XW26}) and can be found online at \url{https://physionet.org/content/mimiciii/1.4/}.

\appendix

\section{{Derivation of importance sampling estimand}}
{We show a derivation of Equation (\ref{eq:vnest})}. Note,
\label{vnaughtIPW}
\vnaughtIPW

\section{Proof that the maximizer of unpenalized value is an indicator}
\label{app:indicator}
{Although Equation (\ref{eq:determinism}) is a known result,\citep{puterman2014markov,Chakraborty2013,jones2022valid}  we prove it here, for completeness, in our notation and for our problem setting.}
\indicator

\section{{Proof of Determinism Corollary}}
{We prove Corollary \ref{coro:deterImpInf}.}
\label{app:deterImpInf}
\determImpliesInf

\section{{Proof of Finite Parameters}}
{We prove Lemma \ref{lemma:finiteBeta0lambda}.}
\label{finiteBeta0lambda}
\proofFinite

\section{The variance of the value estimator}
\label{sigmaV}
\newcommand{\sigmaV}{
We derive a conservative estimator for the {asymptotic} variance of ${\sqrt{n}}V_n(\beta,b_n),$ which we will denote $\sigma^2_V.$   Note that we do not derive an expression for the variance of $J_n=V_n-\lambda||\beta-b_n||.$  We focus on $V_n,$ because we are interested in the value of the policy $\pi_{\beta_{n,\lambda}},$  $V_n(\beta_{n,\lambda},b_n),$ which we can report to the end-user to give a sense of how much the suggested policy might improve reward on average. In contrast, $J_n$  provides the value adjusted by the penalty, and the objective $J_n$  is useful for obtaining  $\beta_{n,\lambda},$ but $J_n$ does not, like $V_n,$ have an interpretation that can be explained to the end-user.

Our estimator for $\sigma^2_V$ is thus
\begin{equation}
\label{eq:sigmaV}
\sigma_n^2(V_n) = \frac{1}{n}\sum_{i=1}^n \left( \frac{\pi_{\beta}(A_{i,0}=a_{i,0}|S_{i,0}=s_{i,0})}{\pi_{b_n}(A_{i,0}=a_{i,0}|S_{i,0}=s_{i,0})} R(S_{i,0},A_{i,0},S_{i,1})-V_n(\beta,b_n)\right)^2.
\end{equation}
%\begin{proof}
%That $V_n(\beta,b_n)\gol \text{N}(V_0,\sigma^2)$ follows from the Central Limit Theorem and the ${\sqrt{n}}-$consistency of $b_n$, which is a maximizer of a generalized linear model likelihood. 

To show how we obtain $\sigma_n^2(V_n),$ recall that $\{S_{i,0},A_{i,0},S_{i,1}\}$ are independent and identically distributed and write,
\begin{align*}
\sigma^2_V&=\text{var}\left(\sqrt{n}V_n(\beta,b_0)\right)\\
&= n\text{var}\left(\frac{1}{n}\sum_{i=1}^n \frac{\pi_{\beta}(A_{i,0}=a_{i,0}|S_{i,0}=s_{i,0})}{\pi_{b_0}(A_{i,0}=a_{i,0}|S_{i,0}=s_{i,0})} R(S_{i,0},A_{i,0},S_{i,1})\right)\\
&= \frac{1}{n}\sum_{i=1}^n \text{var}\left( \frac{\pi_{\beta}(A_{i,0}=a_{i,0}|S_{i,0}=s_{i,0})}{\pi_{b_0}(A_{i,0}=a_{i,0}|S_{i,0}=s_{i,0})} R(S_{i,0},A_{i,0},S_{i,1})\right)\\
&= \text{var}\left( \frac{\pi_{\beta}(A_{1,0}=a_{1,0}|S_{1,0}=s_{1,0})}{\pi_{b_0}(A_{1,0}=a_{1,0}|S_{1,0}=s_{1,0})} R(S_{1,0},A_{1,0},S_{1,1})\right)\\
&= E_{b_0}\left( \frac{\pi_{\beta}(A_{1,0}=a_{1,0}|S_{1,0}=s_{1,0})}{\pi_{b_0}(A_{1,0}=a_{1,0}|S_{1,0}=s_{1,0})} R(S_{1,0},A_{1,0},S_{1,1})-V_0(\beta)\right)^2.
%&= \frac{1}{n}\text{E_{b_0}}\left( \frac{\pi_{\beta}(A_{1,0}=a_{1,0}|S_{1,0}=s_{1,0})}{\pi_{b_0}(A_{1,0}=a_{1,0}|S_{1,0}=s_{1,0})} R(S_{1,0},A_{1,0},S_{1,1})-V_0(\beta,b_0)\right)^2\\
\end{align*}
Hence, we set \[\sigma^2_n(V_n)= \frac{1}{n}\sum_{i=1}^n\left( \frac{\pi_{\beta}(A_{i,0}=a_{i,0}|S_{i,0}=s_{i,0})}{\pi_{b_n}(A_{i,0}=a_{i,0}|S_{i,0}=s_{i,0})} R(S_{i,0},A_{i,0},S_{i,1})-V_n(\beta,b_n)\right)^2.\] 
}
\sigmaV

This is a conservative estimator for the variance as it does not take into account that the behavioral policy is estimated using the data. 

\section{Monte-Carlo {or repeated sample} confidence intervals for the coefficients of the suggested policy}
\label{mcCIbetanlambda}
In the simulations, since we have repeated Monte-Carlo datasets, {and in the real data, since we perform repeated test-train splits,} we compute confidence intervals for $\beta_{n,\lambda}.$  This serves as an indicator of the degree of variability in these estimates, and  these intervals are useful to see in the figures.  Although it is not certain that $\beta_{n,\lambda}$ has a representation as an average, we assume that roughly \[\sqrt{n}(\beta_{n,\lambda}-\beta_{0,\lambda})\gol \text{N}(0,\sigma^2(\beta_{\lambda})),\] where $\sigma^2(\beta_{\lambda})$ is the variance in the limit.
To construct confidence intervals, for $M$ Monte-Carlo datasets {or repeated test-train splits}, we estimate $\sqrt{\sigma^2(\beta_{\lambda})/{n}}$, {the standard error,} with \[\sigma_n(\beta_{n,\lambda})= \sqrt{var_M\left(\beta_{n,\lambda}^{(m)}\right)},\] where $\beta_{n,\lambda}^{(m)}$ is the estimate from Monte-Carlo dataset $m,$ {which can be estimated by computing the empirical variance of the $M$ Monte-Carlo estimators, $\beta_{n,\lambda}^{(1)},\dots,\beta_{n,\lambda}^{(M)}$ }, where the empirical variance of a random estimator $x^{(m)}$ is $var_M(x^{(m)})=\frac{1}{M{-1}}\sum_{m=1}^M (x^{(m)}-\frac{1}{M}\sum_{m=1}^Mx^{(m)})^2$. 
We then construct a $95\%$ confidence interval for $\beta_{0,\lambda}$ {by adding or subtracting the standard error}, \[\beta_{n,\lambda}\pm z_q\sigma_n(\beta_{n,\lambda}),\] where $z_q$ is the $97.5\%$ quantile of a standard normal distribution {}. Although the intervals are not necessarily nominal, they give bounds that reflect the uncertainty of the estimate $\beta_{n,\lambda}.$  {Note that these confidence intervals for the coefficients have different meaning under Monte-Carlo than under repeated sampling of the real data.}
%Note that possibly the variance could be obtained from bootstrap \citep{efron1986bootstrap}.

\section{Scaling}
\label{scaling}
Note that we must scale the state covariates for the penalty.  In particular, we want that for all  $k\neq k'$, our estimates ${\beta}_{n,k}$ and ${\beta}_{n,k'}$ are constrained equally.  Standardization for reinforcement learning has been discussed,\citep{hoffman2011regularized} and we take a similar approach here. For  $ i=1,\dots,n, k=1\dots,K,$ let $\sigma_{n,k}$ be an estimate for the standard deviation of the dimension $k$ of the state.  So that we do not have to rename all states $S_0$ to $\tilde{S}_0,$ in an abuse of notation, let $\tilde{S}_0$ refer to the raw, unscaled states.  Then scale $(\tilde{S}_{i,0})_k$ as $({S}_{i,0})_k={(\tilde{S}_{i,0})_k}/{\sigma_{n,k}}.$   

\section{{Derivation of  simulation estimand}}
\label{app:proofSimBetaInf}
{We provide a derivation of Equation (\ref{eq:sim.application}).}

\simulationBetaInf

\section{Evaluating the  specification of the behavioral policy model in the real data}
\label{app:calc}

To evaluate the specification of the behavioral policy model in Equation (\ref{eq:bpolicy}), we show a calibration curve in Figure \ref{calcurve}.
\begin{figure}
    \centering
\includegraphics[width=0.5\textwidth]{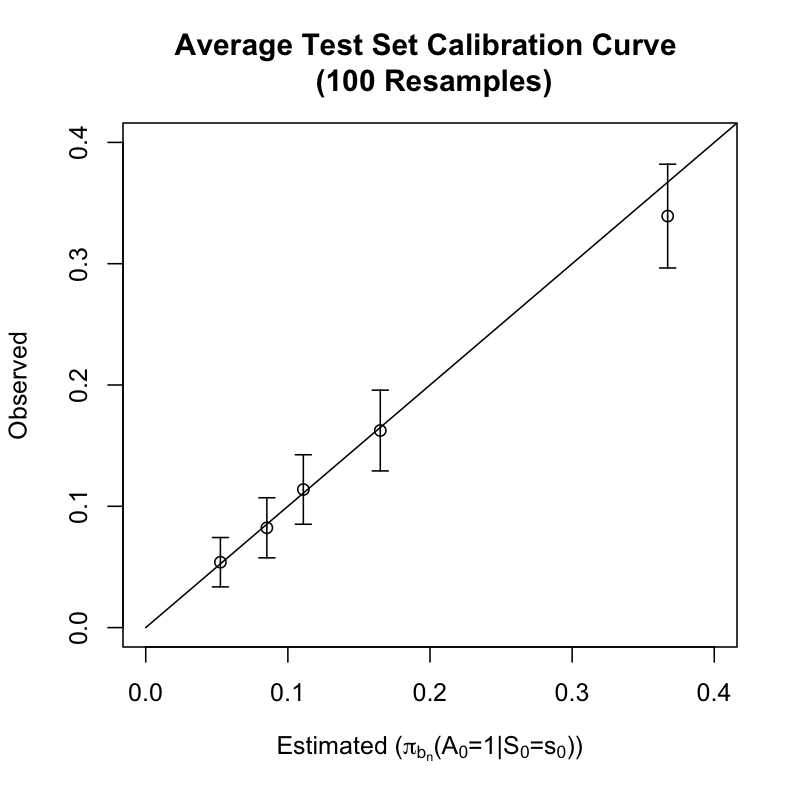}
    \caption{ {{\bf {Real data }calibration curve for the behavioral policy.} We show a calibration curve for the real data, in which the estimated and observed probabilities are compared on held out data. Note that the points are chosen according to the quintiles of the distribution of the predictions. We resample the data, each time training on one half and then generating a calibration curve for the test data, and then we finally average these curves.
    %Note that although the density of the predictions is fairly concentrated, this calibration curve resembles the oracle calibration curve in Figure \ref{truecalcurve} more than it does the pathological calibration curve in Figure \ref{rulebasedcalcurve}, 
    The high concordance of the behavioral estimates, $\pi_{b_n}(A_0=1|S_0=s_0)$, with the observed probabilities serves as evidence that the behavioral policy posited in Equation (\ref{eq:bpolicy}) is reasonable.} 
    }
    \label{calcurve}
\end{figure}

\newpage
\section{Pseudocode}

Note that the code is also provided as part of the submission and will be posted to github.
{
\tiny
\label{pseudocode}
\begin{verbatim}
# Hashtag (#) indicates comments
Input:
    M # Number Monte-Carlo or resamples of real data
    Real dataset if not Monte-Carlo
    Lambda grid for suggested policy
    Lambda grid for behavioral policy
    X # Targeted number standard errors above behavioral value 
    C # Number diverging coefficients
Repeat M times and average: 
    If simulation:
        Generate data and split into test and train
    Else:
        Randomly sample real data and split into test and train
        Fit behavioral policy on training data using cv.glmnet with lambda grid for behavioral policy
        Fit a policy on training data for each lambda, using relative sparsity objective function
        Compute training and test value for each policy
        Compute training and test value variance for each policy
        Select lambda on train data 
        # Useful to have one for each dataset to see distrib
        # Target value X SE above behavior with C nonbehavioral coefficients
Return:
        Set of M suggested policies # E.g., if want to inspect distribution
        Average coefficients over M datasets
        Average train and test value over M datasets
        # Will be used to select final policy on average coefficients and value


# Check the calibration of the behavioral policy
Compute (test set) average calibration curve on real data using M test-train splits
# We do not check in simulation, since we know the true generating policy

Plot distributions of M policy value, probability treatment, and relative sparsity

If real data and calibration is not reasonable:
    Break # Nehavioral policy is not well specified; do not proceed with real data analysis
Else if calibration is reasonable, or in simulation:
    Plot behavioral coefficients as horizontal lines
    Plot the suggested policy coefficients as they vary over lambda (with confidence intervals)
    Plot the train and test value as they vary over lamdba (with confidence intervals)
    Select lambda_bar on average coefficients and average value
    # Target value X SE above behavior with C nonbehavioral coefficients
    # This was done within each dataset, but here it is done on the average, which is less noisy
    Final policy is one indexed by lambda_bar
    Plot lambda_bar on value plot, visualize train value of lambda bar
    Plot lambda_bar on coefficient plot (train data), visualize number of diverging coefficients
    
If train data value and relative sparsity adequate
    Stop: use policy with lambda_bar
    Assess test value of policy with lambda_bar
    # Best estimate of real world value
Else:
    Increase or decrease C, X, gather more data, or reassess reward
    Start from beginning
\end{verbatim}
}

\end{document}